%% file: sample631.tex
\documentclass[twocolumn]{aastex631}

\usepackage{multirow}

\newcommand{\s}{\text{ }}
\newcommand{\g}{\textit{g}}
\newcommand{\h}{\hspace{0.08cm}}

\usepackage{soul}

\begin{document}

\include{LHS_6343_properties}
\include{affiliations}


\title{Revisiting Physical Parameters of the Benchmark Brown Dwarf LHS~6343\,C\\
Through a HST/WFC3 Secondary Eclipse Observation}

\author{William Frost}
\affiliation{\UdeM}
\affiliation{\iREx}

\author[0000-0003-0475-9375]{Loïc Albert}
\affiliation{\UdeM}
\affiliation{\iREx}

\author[0000-0001-5485-4675]{René Doyon}
\affiliation{\UdeM}
\affiliation{\iREx}

\author[0000-0002-2592-9612]{Jonathan Gagné}
\affiliation{\PRTA}
\affiliation{\iREx}

\author[0000-0001-7516-8308]{Benjamin T. Montet}
\affiliation{\UNSW}
\affiliation{\Udash}

\author[0000-0002-2428-9932]{Clémence Fontanive}
\affiliation{\UdeM}
\affiliation{\iREx}

\author[0000-0003-3506-5667]{Étienne Artigau}
\affiliation{\UdeM}
\affiliation{\iREx}

\author{John Asher Johnson}
\affiliation{\HSCA}

\author[0000-0002-5494-3237]{Billy Edwards}
\affiliation{\SRON}

\author[0000-0001-5578-1498]{Björn Benneke}
\affiliation{\UdeM}
\affiliation{\iREx}




\begin{abstract}

The LHS~6343 system consists of a resolved M-dwarf binary with an evolved, negligibly irradiated brown dwarf, LHS~6343\,C, orbiting the primary star. Such brown dwarf eclipsing binaries present rare and unique opportunities to calibrate sub-stellar evolutionary and atmosphere models since mass, radius, temperature and luminosity can be directly measured. We update this brown dwarf's mass ($\sMASSc$) and radius ($\sRADc$) using empirical stellar relations and a Gaia DR3 distance. We use Hubble Space Telescope/WFC3 observations of an LHS~6343\,C secondary eclipse to obtain a NIR emission spectrum, which matches to a spectral type of  $\sSPTc$. 
We combine this spectrum with existing Kepler and Spitzer/IRAC secondary eclipse photometry to perform atmospheric characterization using the ATMO-2020, Sonora-Bobcat and BT-Settl model grids. ATMO-2020 models with strong non-equilibrium chemistry yield the best fit to observations across all modelled bandpasses while predicting physical parameters consistent with Gaia-dependant analogs. BT-Settl predicts values slightly more consistent with such analogs but offers a significantly poorer fit to the WFC3 spectrum. Finally, we obtain a semi-empirical measurement of LHS~6343\,C's apparent luminosity by integrating its observed and modelled spectral energy distribution. Applying knowledge of the system's distance yields a bolometric luminosity of log($L_{\text{bol}}/L_{\sun}$) = $\vLBOLc \pm \eLBOLc$ and, applying the Stefan-Boltzmann law for the known radius, an effective temperature of $\sTEFFc$. We also use the ATMO-2020 and Sonora-Bobcat evolutionary model grids to infer an age for LHS~6343\,C of $\sAGEc$ and $\sAGEcSB$ respectively.

\end{abstract}



\section{\textbf{Introduction}}
\label{sec:Intro}

Brown dwarfs (BDs) are substellar objects that bridge the mass gap between the most massive planets (\mbox{$\sim13\,\,\massJUP$}) and the least massive stars (\mbox{$\sim78.5\,\,\massJUP$}) \citep{Spiegel_2011,Chabrier_2023b}. The upper mass constraint is derived from the hydrogen burning limit, which means brown dwarfs are unable to sustain a long-term energy source (i.e., hydrogen fusion) like main-sequence stars do. They can be massive enough to initiate lower-tier reactions such as deuterium fusion, but these can only last a few tens of Myr at best \citep{Chabrier_2000}, which does not allow for BDs to behave like stars over their lifetime. This lack of an internal energy source means that their luminosity and effective temperature inevitably decrease over time as the BD radiates away the thermal energy brought upon by its creation.

Determining the effect that the cooling trend has on BD physical parameters is essential due to the degeneracy in mass, age, effective temperature and luminosity it causes. For example, the mass-age degeneracy makes it impossible to distinguish between a young, low-mass and an old, high-mass BD using photometric and/or spectroscopic observations alone. Since most reported BDs are isolated field objects, oftentimes luminosities, colours and spectral types are the most that can be directly measured without modelling efforts. Unlike main-sequence stars, BDs do not offer the luxury of constant luminosity which allows for empirical relations between physical parameters, such as the M dwarf mass-luminosity relation of \cite{Mann_2019}.

Consequently, substellar evolution and atmosphere models have been used extensively to infer the properties of BDs. 
Due to their importance, anchoring the predictions of these models to observed BD properties has been a long-standing goal. 
Nearby BD eclipsing binaries (BDEBs) are thus key systems that can be probed using radial velocity, transit and secondary eclipse observations to yield direct measurements of the mass, radius, luminosity and effective temperature of a BD. To serve as adequate benchmarks for field BD characterization, an additional constraint is that BDEBs should be subject to minimal irradiation (and other phenomena, see \citealt{beatty.2018}) from their host star. 

The relatively large population of field BDs is contrasted by the scarcity of reported BDEBs, a phenomenon commonly referred to as the "Brown Dwarf Desert" (e.g., \cite{Grieves_2017}). Indeed, considering the extensive catalogue of over eleven thousand brown dwarfs to date (e.g., \citealt{Carnero_Rosell_2019}), only 37 BDEBs are documented in the literature \citep{Carmichael_2023}. Furthermore, only a subset of these ($\sim 10$) orbit M dwarfs, which allow for higher signal-to-noise observations around empirically calibrated, low-mass main-sequence stars. An even smaller subset possesses precisely measured ages (e.g., \citealt{beatty.2018}), a particularly crucial parameter for young BDs since their luminosities and radii contract quickly below 1 Gyr. 
Other phenomena, such as host irradiation and/or tidal locking of a BDEB (e.g., \citealt{beatty.2014}), can further differentiate their characteristics compared to field BDs. Ultimately, there are very few BDEB systems that can be used as field BD analogs.

A very favourable nearby BDEB system is LHS\,6343. It consists of a resolved ($\sim0.6"$) red dwarf binary whose primary component, LHS~6343\,A, harbours an unresolved transiting and minimally irradiated BD (LHS~6343\,C) orbiting with a period of 12.7 days. This BD was first reported by \cite{Johnson_2011} within the Kepler Telescope Q0-Q1 public data release. The initial ambiguity in assigning the BD primary to stellar component A or B was lifted through a spectroscopic campaign showing that the brighter primary, LHS~6343~A, showed a periodic RV signal. The most recent analyses on this BD are found in two articles by \cite{Montet_2015,Montet_2016}. In their 2015 paper, Keck/HIRES radial velocity observations and the Kepler transit data set are used to obtain direct measurements of the BD's mass and radius: $62.1\pm1.2\,\,\massJUP$ and $0.783\pm0.011\,\,\radiusJUP$, respectively. Furthermore, a slight eccentricity to the BD's orbit is reported, making it unlikely that it is tidally locked to LHS~6343\,A. 

For their 2016 paper, Spitzer/IRAC observations of four secondary eclipses (2 for each of the IRAC-1 and IRAC-2 bandpasses) are used to obtain model-dependant values of the age ($5\pm1$\,Gyr), bolometric luminosity ($-5.16\pm0.04$\,\,log($\lbolSUN$)) and effective temperature ($1130\pm50\,\KELVIN$) of the BD. The models used were the Dartmouth stellar models \citep{Dotter_2008} and the BT-Settl (CIFIST 2011) substellar models \citep{Allard_2012}. Using the Dartmouth models, it is estimated that irradiation from LHS~6343\,A should contribute only $\sim 1\%$ of the total emitted flux of the BD. Finally, both Johnson et al. and Montet et al. contribute valuable information on the dilution effect of LHS~6343\,B on the observed transit and eclipse events of LHS~6343\,C as it orbits LHS~6343\,A. They respectively measure B-A magnitude differences of $\sKsBA$ and $\sKpBA$ for the 2MASS $K_S$ and Kepler $K_P$ bands.\\

The results presented in this paper build upon those reported in the Montet et al. papers, enabling two new semi-empirical measurements: bolometric luminosity and effective temperature. This is accomplished by constructing an emission spectrum of LHS~6343\,C using a single HST/WFC3-G141 secondary eclipse observation and the existing emission photometry from Kepler and Spitzer data. We also derive new measurements for the BD's mass and radius by updating those same parameters for the host star (LHS~6343\,A) using the \cite{Mann_2019} mass-luminosity and \cite{Mann_2015} radius-luminosity empirical relations as well as a Gaia Data Release 3 (DR3) parallax of LHS~6343\,A \citep{Gaia_2016,GaiaDR3_2023}. However, section~\ref{subsec:gaia-distance-comments} touches on certain doubts regarding the validity of this DR3 value and presents a justification for inflating the reported parallax uncertainty to reflect the empirical findings of \cite{El_Badry_2021}. 

Equipped with a broader coverage emission spectrum and a revised set of physical parameters, this work compares the predictions of two sets of self-consistent atmospheric and evolutionary models: ATMO-2020 of \cite{Phillips_2020} and Sonora-Bobcat of \cite{Marley_2021}. The BT-Settl (CIFIST 2011) atmospheric models of \cite{Allard_2012} are also considered, although without corresponding evolutionary models. Two different atmospheric model fits are performed. The first assumes a known distance $d$ based on Gaia DR3. This ultimately fixes the mass $M$ and radius $R$ of LHS~6343\,C, which in turn fixes the surface gravity $GM/R^2$ and the model flux scaling factor $(R/d)^2$ used to interpolate atmospheric model grids. The second approach instead allows the distance to vary uniformly within realistic bounds, which explores model outcomes unconstrained by the Gaia distance and leaves the BD mass, radius, etc. as correlated free parameters. Finally, we compare the measured Gaia-dependant physical parameters of LHS~6343\,C to the predictions of evolutionary models and infer a completely model-dependent age of the BD.

\vspace{0.25cm}

\section{\textbf{Observations and Data Reduction}}
\label{sec:Obs-Reduc}

LHS~6343 was observed with the Wide Field Camera 3 on the Hubble Space Telescope (HST/WFC3) as part of HST cycle 23 GO (PEP-ID 14142; PI L. Albert). These observations are complemented with previous Spitzer/IRAC eclipse observations  \citep{Montet_2016}, Kepler transit and eclipse observations, as well as Keck/HIRES radial velocity observations \citep{Montet_2015}.

\subsection{\textbf{HST/WFC3 Secondary Eclipse Observation}}
\label{subsec:HST-obs}

The LHS~6343\,C secondary eclipse was observed through a single HST/WFC3 visit spanning 7\,h, consisting of 5 telescope orbits separated by \hbox{$\sim$50-minute} gaps in data collection due to Earth occultation. The time series was obtained with the G141 IR grism in spatial scan mode in which the telescope drifts during the exposure such that the source's spectrum is spread out over several pixels perpendicular to the dispersion axis \citep{McCullough_MacKenty_2012}. The scans had a spatial (i.e., vertical) extent of roughly 100 detector pixels, yielding a higher signal-to-noise for a given exposure without saturating the WFC3 detector. To minimize instrumental overheads, both forward and backward scans were performed at a drift scan rate of 0.1~$\arcsec$/\,s.

In total, the HST/WFC3 time series consisted of 102 scan-mode exposures, each integrated over 103 seconds and broken up into 16 sub-exposures. Additionally, each HST orbit's worth of scan-mode exposures is bounded in time by photometric observations in either the F126N ($\times2$), F128N ($\times1$), F130N ($\times1$), F132N ($\times2$), F164N ($\times2$) or F167N ($\times1$) filters. The first HST orbit is preceded by four additional F130N dithered images as well as four stare-mode (non-scanned) spectra. Only the scanned exposures and one photometric non-dithered exposure were used in the reduction pipeline.

\subsection{\textbf{HST/WFC3 Data Reduction}}
\label{subsec:HST-reduc}

The use of custom reduction pipelines is necessary for HST/WFC3 scan-mode spectroscopic observations, as the technique itself started being implemented roughly 2 decades after the telescope's launch \cite{McCullough_MacKenty_2012}. Therefore, the official HST reduction pipelines are insufficient in properly calibrating raw scanned images. We chose to reduce the last 84 raw scan-mode exposures to raw spectroscopic light curves using the ExoTEP data reduction pipeline \citep{Benneke_2019v1,Benneke_2019v2}. The first HST orbit (18 exposures) was discarded for reasons explained in subsection~\ref{subsubsec:sys-mod}. The $60^{\rm th}$ exposure was also discarded as it deviates significantly from the data.

The reduction implemented by ExoTEP follows standard procedure for scan-mode observations \citep{Benneke_2019v1,Benneke_2019v2}. It minimizes the contribution from the sky background by subtracting consecutive non-destructive reads and then co-adding these background-subtracted sub-exposures. It then uses the wavelength-dependent flat-field data provided by STScI to produce flat-fielded images. Bad pixels are replaced by the corresponding value in a normalized row-added flux template.

The WFC3 G141 grism dispersion profile is dependent on the source's spatial ($y$) axis position. Therefore, spatial scanning results in a slightly trapezoidal spectrum instead of a perfectly rectangular one. To correctly capture this effect, ExoTEP integrates over trapezoidal wavelength bins built from lines of constant wavelength obtained from its 2D wavelength solution computed across the detector. During flux integration, it avoids any pre-smoothing and accounts for partial pixel flux along the wavelength binning boundaries, which ensures total flux conservation. It also accounts for small dispersion ($x$) position shifts in each frame to correct for the small drift in a star’s position across the observations.

\vspace{0.25cm}

\section{\textbf{HST/WFC3 Light Curve Analysis}}
\label{sec:Datanal}

This section details the steps and results in going from raw extracted photometry to a calibrated, absolute $F_\lambda$ spectrum for the LHS~6343\,C secondary eclipse in the WFC3 bandpass. 

\subsection{\textbf{Eclipse White-Light Curve Fitting}}
\label{subsec:WLC-fit}

We first fit the extracted raw HST/WFC3 white-light curve (WLC) to a joint instrument systematics and eclipse model. This is accomplished using the \texttt{emcee} python package \citep{Foreman_Mackey_2013} which implements the \cite{Goodman_Weare_2010} Affine Invariant (AI) Markov chain Monte Carlo (MCMC) ensemble sampler. The joint model as a function of time is simply
\begin{equation}
\label{eq-joint-model}
M(t_{v},t_{o})\s =\s M_{ecl}(t_{v})\s\times\s M_{sys}(t_{v},t_{o})
\end{equation}
where $M_{ecl}$ represents the eclipse model and $M_{sys}$ the instrument systematics model. $t_{v}$ represents the time elapsed during the entire HST eclipse observation (i.e. the visit time) and $t_{o}$ represents the time elapsed during a single HST orbit. The WLC has a wavelength coverage of 1.015 to 1.725 $\mu$m.

\subsubsection{Instrument Systematics Model}
\label{subsubsec:sys-mod}

The necessity for splitting time into 2 reference points comes from the nature of the instrument systematics being modelled. Typical HST/WFC3 observations exhibit a linear trend throughout an entire visit as well as showing an exponential trend for each HST orbit. If scan-mode observations are performed, an additional offset in the flux is introduced for backward scans, as the reported flux in these images will be greater than for forward scans. This is due to the detector pixel readout sequence taking slightly more time to reach the starting point of a backward scan compared to that of a forward scan, allowing for slightly more photoelectrons to accumulate on the detector. All 3 systematics behaviours described here are present in the raw photometric time series shown in Fig.~\ref{fig:WLC}. Following a similar methodology to \citet{Benneke_2019v2}, the systematics model is defined as
\begin{equation}
\label{eq-sys-model}
M_{sys}(t_{v},t_{o}) \s=\s \Big( \s s(t_{v}) + v\s t_{v}\s \Big) \times \Big( \s 1-e^{-a\s t_{o}-b}\s \Big)
\end{equation}
where $v$ represents the visit-long slope term and $a$, $b$ represent the rate and amplitude of the orbit-long exponential term. $s(t_{v})$ is set to 1 for forward scans and is left as a free parameter for backward scans. We do not include the first HST orbit in our analysis, as the data show a stronger ramp-like effect compared to subsequent HST orbits, indicative of the instrument still stabilizing. We also exclude the first forward and backward scan of each HST orbit to further eliminate the ramp effect from the data.

\subsubsection{Eclipse Model}
\label{subsubsec:ecl-mod}

The astrophysical eclipse light-curve model $M_{ecl}$ is computed using the \texttt{Batman} package of \cite{Kreidberg_2015}. To get the complete $M_{ecl}$ model, the Batman relative eclipse signal is multiplied by a normalization constant $N$ such that the model's in-eclipse flux can replicate the observed in-eclipse flux (corrected for systematics). This leads to $M_{ecl}(t_v)=N*M_{batman}(t_{v})$ .

\subsubsection{MCMC Likelihood Model}
\label{subsubsec:mcmc-mod}

The log-likelihood function as input to the \texttt{emcee} Ensemble Sampler is simply a log-normal distribution
\begin{equation}
\text{ln}\hspace{0.07cm}\mathcal{L}\s =\s -\sum_{i}\Bigg[ \frac{\Big( D_i - M_i \Big)^2}{2\hspace{0.08cm}{\sigma_{obs}}^2}
+ \text{ln}(\sigma_{obs}) +
\frac{\text{ln}2\pi}{2} \Bigg]
\label{eq-likemike}
\end{equation}
where $D_i=D(t_{v,i})$ and $M_i=M(\h t_{v,i}\h,\h t_{o,i}\h)$ are the observed and modelled data, respectively. Rather than estimating photometric errors for each data point beforehand, we instead leave the scatter in the data $\sigma_{obs}$ as a free parameter in the MCMC analysis. This is standard procedure for ExoTEP photometric outputs; it essentially yields a single standard deviation estimate for all individual light curve data points.

\begin{figure}[!ht]
\plotone{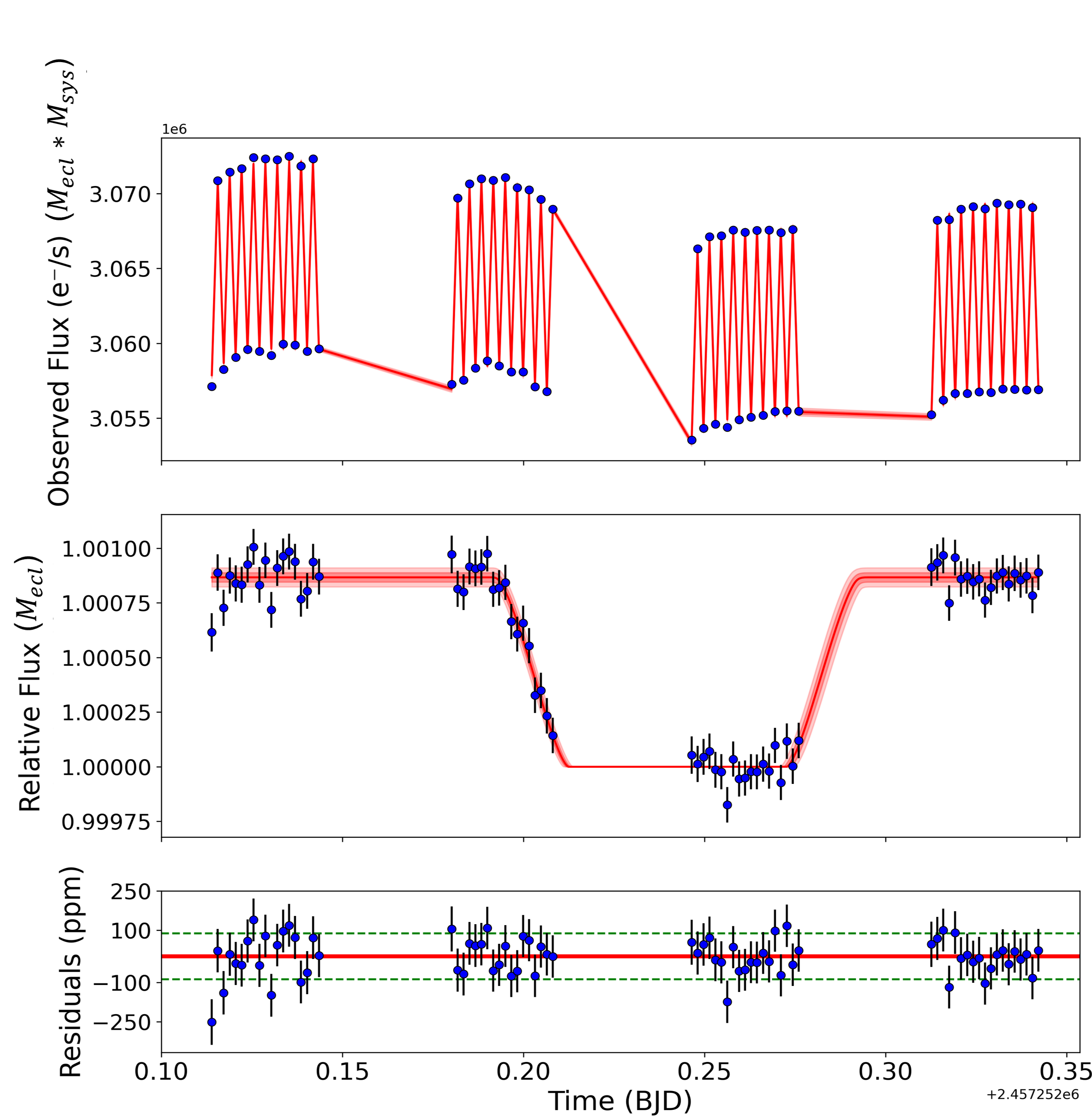}
\caption{White-light curve fit from the analysis of the HST/WFC3 time series of the LHS~6343\,C secondary eclipse. The top plot shows the photoelectron flux as observed by WFC3 in blue. The inferred 1-$\sigma$ error bars are smaller than the size of the symbols. The red line is the median best-fitting model while the shaded regions represent the 1-$\sigma$ and 2-$\sigma$ confidence intervals. The middle plot shows the normalized, systematics-corrected flux with the same approach as the top plot. The bottom plot shows the residuals of the data compared to the best-fitting model with the green dashed lines defining the inferred 1-$\sigma$ envelope.\\
\label{fig:WLC}}
\end{figure}

The sum of the log-likelihood and each fitted parameter's log-prior is used to obtain the joint log-posterior distribution of all astrophysical and instrumental model parameters. For the WLC fit, uniform priors are given to all instrument systematics as well as the eclipse flux ratio $f$, the eclipse conjunction time $t_{ecl}$ and the scatter parameter $\sigma_{obs}$. All known orbital parameters from the literature (Table 2 of \citealt{Montet_2015}) are assigned normally distributed priors about their reported values. The initial values for the parameter chains of the MCMC fit are set to the best-fitting results of a maximum likelihood estimation of Eq. \ref{eq-likemike}. Within emcee, we use 32 walkers for each parameter chain $c_i$ and evaluate convergence by ensuring the chain lengths exceed at least 50 times the autocorrelation time $\tau_i$ computed by \texttt{emcee}. Each parameter chain of 32 walkers is reduced to a singular chain by establishing a burn-in limit of $2\,\tau_{max}$, flattening the 32 walkers and retaining chain realizations at intervals of $0.5\,\tau_{min}$. The main result of the fit is a determination of the eclipse depth in the HST/WFC3-G141 bandpass of $867 \pm 21$\,ppm, marginalized over all WLC parameter chains. The best-fit model to the WLC is shown in Fig.~\ref{fig:WLC}.

\subsection{\textbf{Secondary Eclipse Spectroscopy}}
\label{subsec:SLC-fit}

We use the results from the WLC fitting of section~\ref{subsec:WLC-fit} to inform the priors on a larger number of parameters for the MCMC fit to the spectroscopic light curves (SLCs). Because the HST/WFC3 systematics are independent of wavelength to first-order, all such parameters for each SLC are given Gaussian priors informed by the WLC fit. The same is done for the eclipse conjunction time. The remaining parameters with uniform priors are the flux ratio $f$, the normalization constant $N$ and the scatter parameter $\sigma_{obs}$, as these do vary with wavelength.
We individually fit 27 SLCs between 1.11 and 1.65\,$\mu$m (20\,nm bins), each time performing the same autocorrelation check on the parameter chains as was done for the WLC fit. The results of these fits are visualized in Fig.~\ref{fig:SLC} of the appendix and reported in the ``Relative Flux" column of Table~{\ref{tab:emspec}.

\subsection{\textbf{LHS~6343\,C Emission Spectrum}}
\label{subsec:emspec}

Flux-calibration of the relative emission spectrum measured in section~\ref{subsec:SLC-fit} was performed by multiplying it with the observed LHS~6343\,A+B flux. This combined stellar flux of the system is contained within the results of the SLC fits of section~\ref{subsec:SLC-fit}, as the normalization constant $N$ for each spectroscopic bin retrieves the parameter directly. The stellar photo-electron count rate is then multiplied with the appropriate sensitivity file for HST/WFC3-G141 supplied by STScI to obtain the flux-calibrated spectrum of LHS~6343\,A+B in units of erg~s$^{-1}$~cm$^{-2}$~$\mu$m$^{-1}$. To further validate this spectrum, we anchor it to existing $J$-band photometry from \cite{Johnson_2011} of the whole system whose flux is dominated by the combined light of the two M dwarfs. We find that the $J$-band magnitude of our measured LHS~6343\,A+B spectrum is indistinguishable (within $1\sigma$) of the value reported by \citet{Johnson_2011}.

Finally, we multiply this in-eclipse spectrum by the relative emission spectrum of LHS~6343\,C, yielding a flux-calibrated spectrum of the BD in the HST/WFC3-G141 bandpass. This spectrum is listed in Table~\ref{tab:emspec} and shown in Figures~\ref{fig:spec_types}, \ref{fig:atmos-fit}, \ref{fig:atmos-fit-3} and \ref{fig:atmos-fit-4}. To summarize, our resulting LHS~6343\,C spectrum can be expressed as
\begin{equation}
    F_C = \frac{F_{ABC} - F_{AB}}{F_{AB}} \times \frac{F_{AB}}{T_{WFC3}} \s\s ,
\label{eq:emspec}
\end{equation}
where $F_{ABC}$ is the flux of the entire LHS 6343 system (out-of-eclipse) observed using the HST/WFC3 detector and the G141 IR grism. $F_{AB}$ is the flux of the A and B components (in-eclipse). $T_{WFC3}$ is the transmission/sensitivity profile for WFC3-G141. The first term of Equation~\ref{eq:emspec} effectively represents the emission spectrum of LHS~6343\,C relative to that of LHS~6343\,A+B, and the second term is the absolute flux-calibrated A+B spectrum.

\begin{deluxetable}{lccc}[!b]
\tabletypesize{\scriptsize}
\tablewidth{0pt} 
\tablecaption{ IR Emission Spectrum of LHS~6343\,C \label{tab:emspec}}
\tablehead{
    \colhead{Instrument} & \colhead{Wavelength} & 
    \colhead{Relative Flux} & \colhead{Apparent $F_{\lambda}$} \\
    \colhead{} & \colhead{$\mu$m} & \colhead{ppm} &
    \colhead{$\frac{erg}{s\text{ }cm^2\text{ }\mu m}*10^{-13}$}
}
\startdata
Kepler    &  0.42-0.91  &  25 ($\pm7$)      &  0.024 ($\pm$0.0068)  \\
HST/WFC3  &  1.11-1.13  &  503.6{ } ($\pm$93.7)  &  2.69 ($\pm$0.50)  \\
{      }  &  1.13-1.15  &  418.4{ } ($\pm$115.8)  &  2.14 ($\pm$0.59)  \\
{      }  &  1.15-1.17  &  627.5{ } ($\pm$81.8)  &  3.22 ($\pm$0.42)  \\
{      }  &  1.17-1.19  &  802.7{ } ($\pm$81.8)  &  3.98 ($\pm$0.41)  \\
{      }  &  1.19-1.21  &  979.1{ } ($\pm$82.3)  &  4.74 ($\pm$0.40)  \\
{      }  &  1.21-1.23  &  1261.7 ($\pm$77.7)  &  5.99 ($\pm$0.37)  \\
{      }  &  1.23-1.25  &  1190.4 ($\pm$85.1)  &  5.49 ($\pm$0.39)  \\
{      }  &  1.25-1.27  &  1402.6 ($\pm$88.8)  &  6.37 ($\pm$0.40)  \\
{      }  &  1.27-1.29  &  1502.8 ($\pm$71.3)  &  6.72 ($\pm$0.32)  \\
{      }  &  1.29-1.31  &  1409.0 ($\pm$88.1)  &  6.17 ($\pm$0.39)  \\
{      }  &  1.31-1.33  &  1340.7 ($\pm$89.6)  &  5.59 ($\pm$0.37)  \\
{      }  &  1.33-1.35  &  622.0{ } ($\pm$83.2)  &  2.48 ($\pm$0.33)  \\
{      }  &  1.35-1.37  &  359.3{ } ($\pm$78.8)  &  1.38 ($\pm$0.30)  \\
{      }  &  1.37-1.39  &  402.3{ } ($\pm$89.4)  &  1.52 ($\pm$0.34)  \\
{      }  &  1.39-1.41  &  405.1{ } ($\pm$84.9)  &  1.49 ($\pm$0.31)  \\
{      }  &  1.41-1.43  &  261.7{ } ($\pm$91.0)  &  0.94 ($\pm$0.33)  \\
{      }  &  1.43-1.45  &  392.8{ } ($\pm$86.7)  &  1.37 ($\pm$0.30)  \\
{      }  &  1.45-1.47  &  510.4{ } ($\pm$76.5)  &  1.74 ($\pm$0.26)  \\
{      }  &  1.47-1.49  &  619.0{ } ($\pm$98.2)  &  2.09 ($\pm$0.33)  \\
{      }  &  1.49-1.51  &  729.7{ } ($\pm$94.9)  &  2.42 ($\pm$0.32)  \\
{      }  &  1.51-1.53  &  1007.8 ($\pm$76.2)  &  3.30 ($\pm$0.25)  \\
{      }  &  1.53-1.55  &  1176.3 ($\pm$96.1)  &  3.79 ($\pm$0.31)  \\
{      }  &  1.55-1.57  &  1379.2 ($\pm$89.5)  &  4.45 ($\pm$0.29)  \\
{      }  &  1.57-1.59  &  1564.7 ($\pm$87.8)  &  4.96 ($\pm$0.28)  \\
{      }  &  1.59-1.61  &  1496.6 ($\pm$93.3)  &  4.66 ($\pm$0.29)  \\
{      }  &  1.61-1.63  &  1277.2 ($\pm$98.4)  &  3.89 ($\pm$0.30)  \\
{      }  &  1.63-1.65  &  1220.4 ($\pm$86.5)  &  3.51 ($\pm$0.25)  \\
Spitzer/IRAC-1  &  3.13-3.96  &  1060 ($\pm$210)  &  0.268 ($\pm$0.062)  \\
Spitzer/IRAC-2  &  3.92-5.06  &  2090 ($\pm$80)  &  0.232 ($\pm$0.024)  \\
\enddata
\tablecomments{Kepler and Spitzer relative fluxes are taken directly from \cite{Montet_2015,Montet_2016}. The Kepler absolute flux is calculated using the $K_P$ apparent flux of LHS 6343 reported by \cite{Johnson_2011}, the $K_P$ B-A relative magnitude (empirical prior) and transit depth reported by \cite{Montet_2015}. The Spitzer absolute fluxes are calculated by converting the absolute magnitudes for the BD reported by \cite{Montet_2016} to apparent fluxes using Montet's distance measure.}
\end{deluxetable}

\section{\textbf{Brown Dwarf Modelling}}
\label{sec:BD-mod}
\vspace{0.25cm}

Since the last detailed analyses of the LHS~6343 system by \cite{Montet_2015, Montet_2016}, a Gaia DR3 parallax has become available and the stellar mass-luminosity relation of \cite{Mann_2019} was published. Updating the LHS~6343 system properties is our first step in deriving physical parameters for LHS~6343\,C. The system's distance directly impacts the luminosities of the LHS 6343 members, which in turn affects the mass and radius of the primary M dwarf of the system (LHS~6343\,A) determined using empirical stellar relations, and thus the derived BD mass and radius from radial velocity and transit observations. Ultimately, assumptions on the distance also impact atmospheric and evolutionary modelling results, as these use the physical parameters of LHS~6343\,C as direct inputs.

\subsection{\textbf{Distance to LHS~6343}}
\label{subsec:gaia-distance-comments}

The Gaia DR3 release provides the first parallax measurement of the LHS~6343 system, corresponding to a distance of $\sDISTg$. In principle, a Gaia distance would lift the dependency on stellar models that limited previous studies in deriving model-free BD radius and mass. The binary nature of the system, however, complicates the matter. The transiting BD, LHS~6343~C, orbits the primary component of a binary system whose two stellar components are resolved with a projected separation of 0.55" \citep{Johnson_2011}.

What is problematic is that LHS~6343 does not appear in the two-body orbit table of DR3 and, instead, is fit using a 5-parameter astrometric model, assuming a single star. The \textit{Renormalized Unit Weight Error} (RUWE) of the source listed as LHS~6343~A is 12.4 compared to a desired RUWE$\,\leq1.4$ expected for a good fit to a single-star model. The single-star astrometric fitting process is used despite two other flags clearly pointing towards LHS~6343 as being resolved by Gaia: \textit{ipd\_frac\_multi\_peak} of 80 and an \textit{ipd\_gof\_harmonic\_amplitude} of $\sim0.115$ are sensitive to resolved binaries \citep{Lindegren_2021}. Also, a second source is resolved by Gaia with a $~0.7\,\arcsec$ apparent separation and $G=13.3$, which is likely to be LHS~6343\,B.
 
How sound is the Gaia distance then? Using a test sample of eclipsing binaries with known distances, \cite{stassun.2021} explain that large RUWE values generally come from an unseen tertiary component (in our case, the binary LHS~6343~B) and sees cases where the parallax is off by 50\% for such large RUWE. That represents an extreme case where Gaia does not resolve the tertiary component. For LHS~6343, the stellar components \emph{are} resolved so the parallax error is likely less than 50\%. Assuming an error of half that of the eclipsing binaries, $\approx$25\%, this would translate to a loosely constrained distance of $36^{+12}_{-7}$~pc.

According to \cite{Lindegren_2021}, the parallax standard uncertainties given in DR3 have been adjusted to take into account the excess noise diagnosed by the large RUWE. However, \cite{El_Badry_2021} find that Gaia parallax uncertainties can be underestimated by as much as a factor of 2-3 for resolved objects with similar characteristics as LHS~6343\,A ($G\approx13$, projected separation $\leq2\arcsec$, RUWE$\,\geq1.4$). Since LHS~6343\,A is seemingly resolved by Gaia, we inflate the reported parallax error by a factor of 3 to account for the findings of \cite{El_Badry_2021}, yielding a distance measurement of $\sDIST$.

Ultimately, the parallax concern will only get resolved with a future Gaia release adopting a binary astrometric model. Lacking any better measurement, we adopt this conservative Gaia distance to update the physical parameters of the system (Sec.~\ref{subsec:update-stella}) as well as for fitting model atmospheres using a fixed distance (Sec.~\ref{subsubsec:constrainModFit}). The impact of applying no prior knowledge on distance is presented in section~\ref{subsubsec:unconstrainModFit}.

\subsection{\textbf{Updated Stellar Parameters for LHS 6343}} 
\label{subsec:update-stella}

Partly because they lacked a parallax measurement, the analysis carried by \cite{Montet_2015} implemented two different techniques to set priors on the LHS 6343 A mass during a joint fit to the radial velocity and transit observations for LHS~6343\,C. One was dubbed an ``empirical'' prior, as the stellar mass value was informed strictly from the empirical mass-radius relation of \cite{Boyajian_2012}. The other method used a ``model'' prior, where stellar mass was now informed from a near-IR spectroscopic analysis combined with the predictions of the Dartmouth stellar evolution models. The empirical prior approach could yield only the mass and radius of the primary M dwarf, while the model-dependent approach could yield those for the secondary M dwarf as well. The model prior approach also allowed them to get estimates of effective temperature for both stars, and the inferred stellar absolute magnitudes were used to obtain a distance of $32.7\pm1.3$\,\,pc to the system. They also obtained a metallicity value for the system from NIR spectroscopy, which we adopt for our analysis as well.

Having access to a distance measurement from Gaia DR3, we opt instead to use empirical stellar relations to measure stellar masses, radii and effective temperatures. To obtain these new values, we first convert the resolved $K_S$ apparent magnitudes of both stars reported by \cite{Johnson_2011} to absolute magnitudes using the Gaia distance of $\sDIST$. These stellar magnitudes are then converted to masses and radii using the \cite{Mann_2019} mass-luminosity and \cite{Mann_2015} radius-luminosity empirical relations for M dwarfs. Effective temperatures are determined by inverting the radius-temperature relation detailed in \cite{Mann_2015}.

To measure the mass of LHS~6343\,C, we make use of the primary M dwarf mass $M_A$, the primary star's radial velocity semi-amplitude $K_{\text{RV}}$ and relevant BD orbital parameters reported in Table 2 of \cite{Montet_2015} to solve for the BD's mass $M_C$ in Equation 2.27 of \cite{ExoP-Handbook}:
\begin{equation}
K_{\text{RV}} \h=\h \Big( \frac{2\pi G}{P} \Big)^{1/3} \h \frac{M_C\h\h\text{sin}\h i}{(M_A +M_C)^{2/3}} \h\h\h \frac{1}{(1-e^2)^{1/2}}\s\s \textbf{.}
\label{eq:radvel}
\end{equation}
To update the BD radius, we simply multiply the primary star's radius with the reported Kepler transit radius ratio of \cite{Montet_2015}. Updated masses and radii for all LHS~6343 members are reported in Table~\ref{tab:updatedMR}. Ultimately, our BD radius measurement remains consistent with the value reported by \cite{Montet_2015}, albeit with a $3-4$ times larger uncertainty mainly due to the propagation of the uncertainty in our adopted Gaia distance and the empirical stellar relations used. In contrast, our mass measurement is consistent well within 1-$\sigma$ of Montet's model-prior value, while it borders the 1-$\sigma$ envelope of the empirical-prior value.

In Table~\ref{tab:updated-pars-error-breakdown}, we show the importance of various error sources that contribute significantly to the uncertainty of our BD mass and radius measurements, which are obtained via Monte Carlo propagation. The note for Table~\ref{tab:updated-pars-error-breakdown} further explains the propagation scheme used. The uncertainty of the flux ratio between LHS~6343\,A \& B in the Kepler (\nKpBA\s$=\sKpBA$) and 2MASS~$K$ (\nKsBA~$=~\sKsBA$) bandpasses is reflected in the uncertainties reported for the Kepler transit radius ratio \citep{Montet_2015} and the primary M dwarf's $K$-band magnitude \citep{Johnson_2011}. For both radius and mass, the limiting source of error is due to the inflated uncertainty of the distance measurement. Stellar empirical relations follow in magnitude and would have undoubtedly been the primary contributor had a Gaia parallax measurement of good quality been available. 

\begin{deluxetable}{cccc}[!t]
\tabletypesize{\scriptsize}
\tablecaption{Updated LHS 6343 radii \& masses compared to the \cite{Montet_2015} (M15) reported values. \label{tab:updatedMR}}
\tablehead{
    \colhead{Parameter} & \colhead{This Work} & \colhead{M15 Empirical Prior} & \colhead{M15 Model Prior}
}
\startdata
{}&{}&{}&{} \\[-7pt]
$M_A$ [$\uMASSa$] & $\vMASSa\pm\eMASSa$ & $0.381\pm0.019$ & $0.358\pm0.011$ \\
$M_B$ [$\uMASSb$] & $\vMASSb\pm\eMASSb$ & - & $0.292\pm0.013$ \\
$R_A$ [$\uRADa$] & $\vRADa\pm\eRADa$ & $0.380\pm0.007$ & $0.373\pm0.005$ \\
$R_B$ [$\uRADb$] & $\vRADb\pm\eRADb$ & - & $0.394\pm0.012$ \\
$M_C$ [$\uMASSc$] & $\vMASSc\pm\eMASSc$ & $64.6\pm2.1$ & $62.1\pm1.2$ \\
$R_C$ [$\uRADc$] & $\vRADc\pm\eRADc$ & $0.798\pm0.014$ & $0.783\pm0.011$ \\[4pt]
\enddata
\tablecomments{To obtain our mass and radius values, this work used as inputs a Gaia DR3 distance, the LHS 6343 A \& B 2MASS $K_S$ apparent magnitudes reported in Table 1 of \cite{Johnson_2011}, the orbital parameters (period, inclination, eccentricity, radius ratio and radial velocity semi-amplitude) reported in Table 2 of M15 as well as the stellar empirical relations of \cite{Mann_2015} and \cite{Mann_2019}.
}\end{deluxetable}

\begin{deluxetable}{lcc}[!t]
\tablecaption{Individual importance of variance contributors to the LHS~6343\,C mass and radius relative to total variance. \label{tab:updated-pars-error-breakdown}}
\tablehead{
    \colhead{Contributor} & \colhead{Mass ($\massJUP$)} & \colhead{Radius ($\radiusJUP$)} \\
    \colhead{} & \colhead{$\vMASSc\pm\eMASSc$} & \colhead{$\vRADc\pm\eRADc$}
}
\startdata
LHS 6343 distance$^{(a)}$                   & 68.6\% & 50.0\% \\
$K_{\text{S,A}}$$^{(b)}$ (app. magnitude)   & 14.8\% & 10.7\% \\
LHS~6343\,AC $R_{BD}/R_*$$^{(c)}$           & ---    & 11.4\% \\
\cite{Mann_2019} MLR                        & 16.6\%   & ---    \\
\cite{Mann_2015} RLR                        & ---    & 27.9\% \\
\enddata
\tablecomments{\hspace{0.2cm}a) Gaia DR3. b) Table 1 of \cite{Johnson_2011}.\\ c) Table 2 of \cite{Montet_2015}. The importance of each contributor to the total (i.e. reported) $1\sigma$ confidence for the mass and/or radius is obtained by performing a Monte Carlo uncertainty estimation where only the standard deviation of that contributor is propagated. The resulting single-contributor 1-$\sigma$ variance is then compared (as a percentage) to the one resulting from the Monte Carlo estimation where the errors of all contributors were propagated.}
\end{deluxetable}

\vspace{-0.7cm}
\subsection{\textbf{Spectral Classification}}
\label{subsec:sptype}

\begin{figure*}[!ht]
\centering
\includegraphics[scale=0.73]{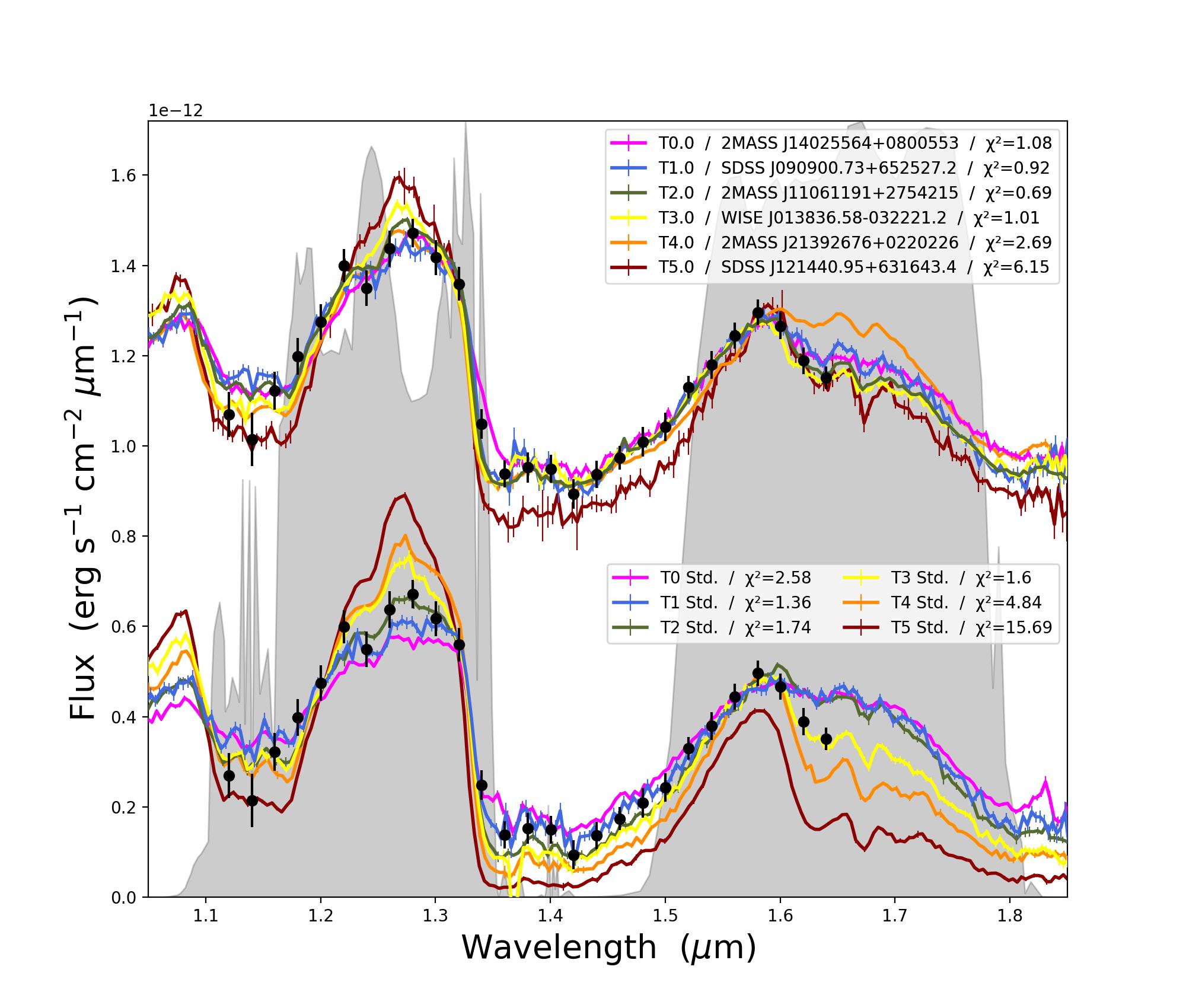}
\caption{HST/WFC3 secondary eclipse emission spectrum of LHS~6343\,C (black) and its spectral typing using spectral standards (bottom) and templates (top) found in the \texttt{SPLAT} python package. The best-matching standards/templates for types T0 to T5 are tested with their associated reduced $\chi^2$. The transmission profiles corresponding to the 2MASS $J$ and $H$ bands are shown in grey, each scaled such that their peaks correspond to the maximum vertical extent of the figure.\\\text{}
\label{fig:spec_types}}
\end{figure*}

We determine a near-IR spectral type for LHS~6343\,C of $\sSPTc$ by matching its observed HST/WFC3 spectrum to the spectral standards (average of several BD spectra with similar spectral type) and templates (individual brown dwarfs) contained within the \texttt{SPLAT} python package of \cite{burgasser2017spex}. Fig.~\ref{fig:spec_types} shows the best-matching spectral standards and templates from T0 to T5. 
Both spectroscopic standards and templates converge towards similar spectral types: T1 for the best-matching standard, and T2 for the best-matching template. The individual templates yield better reduced $\chi^2$ values than the corresponding standards, perhaps reflecting a real diversity within each type. The best-matching template is 2MASS J11061197+2754215, whose spectrum is classified as a T2 within \texttt{SPLAT}, although the source could be a tight-orbit or line-of-sight T0+T4 binary according to \cite{Burgasser_2010}.

One of the most important physical parameters determining spectral type is effective temperature. Table~\ref{tab:spec-type-to-teff} shows the results of applying the various empirical relations found within the \texttt{SPLAT} package that convert the $\sSPTc$ spectral type to an effective temperature. These results show a large dispersion ranging between 1200 and 1450\,K. This is because LHS~6343\,C falls in the L/T transition which is characterized by a large spectral type scatter (e.g., see Fig.~\ref{fig:CMD}), making its treatment vary between empirical relations. As a comparison, in section~\ref{subsubsec:direct-meas}, the $T_{\text{eff}}$ of LHS~6343\,C is directly determined using atmospheric models and the Stefan-Boltzmann law.

\begin{deluxetable}{lcc}[!ht]
\tablewidth{0pt} 
\tablecaption{Converting the LHS~6343\,C Spectral Type (T1.5$\pm$1) to an Effective Temperature  \label{tab:spec-type-to-teff}}
\tablehead{ \colhead{Empirical Relation} &{}& \colhead{$T_{\text{eff}}$ (K)} }
\startdata
\cite{Golimowski_2004}&\s\s\s\s\s\s\s\s\s\s&$1446\pm 126$ \\
\cite{Looper_2008}&{            }&$1375\pm 90$ \\
\cite{Stephens_2009}&{            }&$1208\pm 102$ \\
\cite{Marocco_2013}&{            }&$1351\pm 140$ \\
\cite{filippazzo_2015}&{            }&$1192\pm 117$ 
\enddata
\tablecomments{Conversions obtained using the empirical relations found within the \texttt{SPLAT} python package developed by \cite{burgasser2017spex}.}
\end{deluxetable}

\subsection{\textbf{Color-Magnitude Diagram}}
\label{subsec:CMdiag}

The HST/WFC3 spectrum covers the $J$-band filter and spans a large fraction of the $H$-band. LHS~6343\,C can therefore be positioned in a $M_J$ vs. $J$\,$-$\,$H$ color-magnitude diagram (CMD) for the first time in the NIR where the emergent flux peaks (see Fig.~\ref{fig:CMD}). Because the HST/WFC3 spectrum of LHS~6343\,C does not extend over the full $H$-band wavelength range (which prevents performing synthetic photometry), the photometry of the best-matching template, 2MASS J11061197+2754215, is instead adopted as an estimate for the $J$\,$-$\,$H$ color: $(J$\,$-$\,$H)_{MKO} = 0.76\pm 0.06$ \citep{manjavacas.2013}. The apparent $J$-band magnitude for LHS~6343\,C is measured by spectral synthesis of our WFC3 spectrum at a value of $J=17.632\pm0.025$ (MKO). The Gaia distance corresponds to a distance modulus of $2.76\pm0.11$~mag. Therefore, the absolute magnitude is $M_J=\sKsC$.

LHS~6343\,C lies in the L-T transition, as its spectral type of T1.5$\pm1$ and its position in a CMD are consistent with T0-T2 subtypes, albeit on the faint end of that population. Given that a large spread of magnitude exists for that population ($\pm1$~mag), we can not conclude whether that J-band magnitude is a sign of an underestimated distance to the system or if LHS~6343~C is simply on the faint end of the population. This question is discussed in sections~\ref{subsec:alt-d} and \ref{subsec:viewingangle}.

\begin{figure}[!t]
    \centering
    \includegraphics[width=0.45\textwidth]{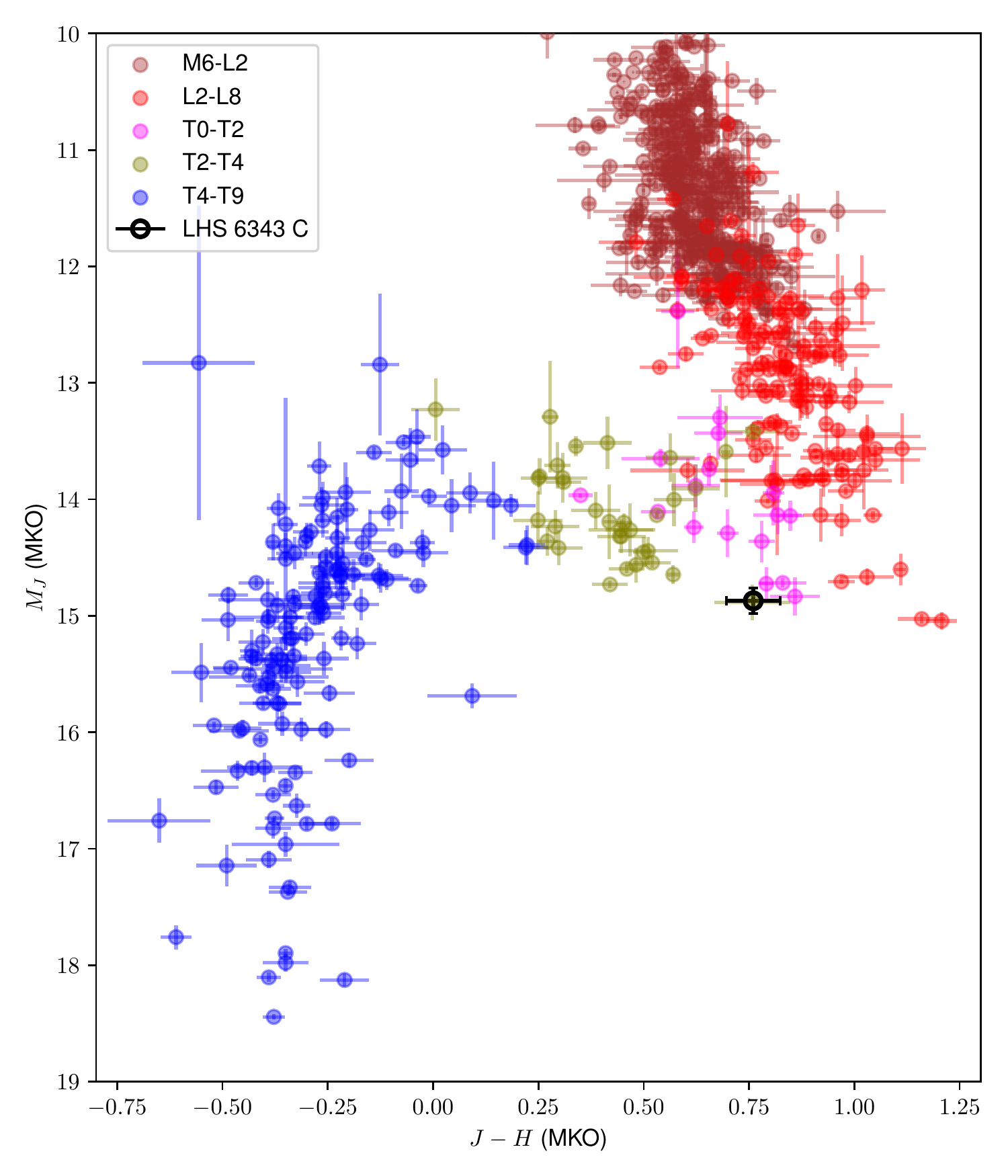}
    \caption{Position of LHS~6343\,C in a $M_J$ vs. $J$\,$-$\,$H$ color-magnitude diagram generated from a compilation of field objects within Jonathan Gagné\hspace{0.04cm}'s MOCA database (\url{mocadb.ca}, private version). This brown dwarf lies in the L/T transition and, for its spectral type of $\sSPTc$, is slightly less luminous than other brown dwarfs of its category (T0-T2), given the Gaia distance measurement.
    }
    \label{fig:CMD}
\end{figure}

\vspace{0.1cm}
\subsection{\textbf{Spectral and SED Fitting}}
\vspace{0.1cm}

\subsubsection{\textbf{Substellar Model Selection}
\label{subsubsec:model-select}}

We use some of the most recent brown dwarf model grids found in the literature: the ATMO-2020 models from \cite{Phillips_2020} and the Sonora-Bobcat models from \cite{Marley_2021}. Both are self-consistent atmospheric and evolutionary grids, meaning that their atmosphere models are used as boundary conditions in their evolution models. ATMO-2020 offers rainout chemical equilibrium (CEQ) and non-equilibrium (NEQ-weak, NEQ-strong) atmospheres for solar metallicities. The non-equilibrium chemistry models incorporate a vertical mixing parameter $K_{\text{zz}}$ which decreases log-linearly as a function of surface gravity. On the other hand, the Sonora-Bobcat models remain at chemical equilibrium but instead vary metallicity.

Both atmospheric grids vary the effective temperature $T_{\text{eff}}$ and surface gravity $\g$. ATMO-2020 has separate $T_{\text{eff}}-\text{log}(\g)$ grids for chemical equilibrium (CEQ), ``weak" non-equilibrium (NEQ-weak) and ``strong" non-equilibrium (NEQ-strong). Sonora-Bobcat has the same structure but for metallicities $[\,{\rm Fe/H}\,] = [-0.5,0,0.5]$. An additional C/O parameter is modelled, but their published grid is too sparse for meaningful interpolation outside of solar (\hspace{0.04cm}$[\text{C/O}]_\odot=1$). Therefore, it is fixed at unity for our Sonora-Bobcat analysis.

With LHS~6343\,C being an L/T transition BD, we also include a model grid which implements cloudy atmospheres to model such BDs: The BT-Settl models of \citep{Allard_2012} with the \cite{Caffau_2011} CIFIST solar chemical abundances. This grid essentially varies $T_{\text{eff}}$, $g$ and metallicity, but this third parameter does not have non-solar values with a sufficiently high surface gravity (log($g$)$\sim5.5$) to represent LHS~6343\,C. Therefore, metallicity is fixed to solar for BT-Settl models in this analysis. This work could not access the complementary evolutionary model grids for BT-Settl. The \cite{Baraffe_2015} models could have proven a valid substitute, however, the grid this work had access to (supplied by the \texttt{SPLAT} package of \cite{burgasser2017spex}) did not extend far enough in time for masses similar to LHS~6343\,C. Therefore, BT-Settl does not have an associated evolutionary grid for our analysis.

\subsubsection{\textbf{Fitting Methodology}
\label{subsec:fit-method}}

We interpolate the ATMO-2020, Sonora-Bobcat and BT-Settl atmospheric model grids to infer a spectral energy distribution (SED) that best fits the observed emission spectrum in the Kepler, HST and Spitzer bandpasses reported in Table~\ref{tab:emspec}. At the highest level of abstraction, the model grids used in this work consider 2-3 independent parameters for atmospheric characterization: the effective temperature $T_{\text{eff}}$ and the surface gravity $\g$ (+ metallicity for Sonora-Bobcat). The interpolation of the emission spectra within a grid is performed first in log-gravity space using a cubic spline, then linearly interpolated in log-$T_{\text{eff}}$ space. For the Sonora-Bobcat models, the contribution of metallicity is derived last using linear interpolation. Finally, the surface flux of atmospheric models must be scaled to match the observed flux of the BD. 

The limits imposed on the exploration of the $T_{\text{eff}}$ parameter space are dependent on the model grid considered. The most restrictive lower limit on $T_{\text{eff}}$ is imposed by the BT-Settl models available to us at $900\,K$, while the upper limit is imposed by the ATMO-2020 NEQ grids at $1800\,K$. Thus, the uniform prior for $T_{\text{eff}}$ across all models is defined using these values. The bounds for exploring surface gravity across all models, regardless of the choice of prior, are set between log\,$\g = [\,2.5\,,\,6.0\,]$. Similar bounds for flux scaling and metallicity are set at $[\,0\,,\,1\,]$ and $[\,-0.5\,,\,0.5\,]$ respectively.

Within this analysis, both the BD mass $M$ and radius $R$ are a function of the distance $d$ to LHS 6343, empirical stellar relations and other parameters as defined in section~\ref{subsec:update-stella}. This effectively makes both the surface gravity $\g=GM/R^2$ and the flux scale factor $(R/d)^2$ correlated functions of $d$, as the other inputs to the empirical relations and equations of section~\ref{subsec:update-stella} are known. Thus, our atmospheric modelling of the LHS 6343 C spectrum adopts 8 base parameters at the lowest level of abstraction:  $T_{\text{eff}}$, distance, the apparent magnitude and radial velocity semi-amplitude of LHS~6343\,A, and the orbital period, inclination, eccentricity and transit depth of LHS~6343\,C. As this base parameter space is explored, instances of the BD mass and radius, and thus surface gravity and flux scaling, are calculated from all base parameters except $T_{\text{eff}}$ to generate interpolated atmospheric models. For Sonora-Bobcat models, metallicity is also a fitted parameter. 

The goodness-of-fit $G_k$ of a model spectrum $k$ is determined by minimizing $G_k$ within a weighted, least-squares framework for simultaneously fitting to photometry and spectroscopy. It is taken from analyses by \cite{Cushing_2008} and \cite{Naud_2014}, and is governed by Equation~\ref{eq:Gk}. In this framework, $D_i$ represents the observed photometry or spectra and $\sigma_i$ the uncertainty in $D_i$. $M_{k,i}$ are the photometry or spectra obtained via interpolation of the model grids. 

\begin{equation}
G_k = \frac{1}{\sum_i W_i}\times\sum_i W_i \h \Bigg(\frac{ D_i - M_{k,i} }{\sigma_{i}}\Bigg)^2 \s\s 
\label{eq:Gk}
\end{equation}

Initial attempts at fitting models to the SED using $\chi^2$ minimization showed that the HST spectrum (27 data points) has a disproportionate influence on results compared to Kepler and Spitzer photometry (3 data points): a bias that persisted despite the photometry's smaller uncertainties. This is because model errors rather than observation uncertainties dominate the $\chi^2$ budget. In other words, the assumption that the data samples a $\chi^2$ distribution does not hold. In that regime, the HST spectrum outweighs the 3 photometric data points. To mitigate this problem, we incorporate an additional weighting $W_i$ for each data point, where $W_i=\Delta\lambda_i$ corresponds to the wavelength coverage of each $D_i$. This effectively puts more emphasis on correctly reproducing the photometry compared to a $\chi^2$ framework. However, such a weighting scheme remains a choice of the authors rather than being intrinsically better than other weighting choices. We also decide to divide the minimization term of Eq.~\ref{eq:Gk} with the sum of the weights $W_i$ to facilitate the comparison of model performance between different combinations of observational data.



As is explained in section~\ref{subsec:gaia-distance-comments}, the binary nature of LHS~6343\,A\,\&\,B could have introduced systematic errors and/or misrepresented the uncertainty in the parallax measurement reported by Gaia DR3. Therefore, we perform fits to model atmospheres using two approaches to handle distance. First, we set the distance at the fixed value informed by Gaia DR3, i.e. $\sDIST$. Second, we allow the distance to vary freely with a large, but realistic, uniform prior. Thus, comparing the results of the two approaches becomes a simultaneous test of the model predictions and of our knowledge of the distance. In both cases, the stellar empirical relations of \citep{Mann_2015} and \cite{Mann_2019} constrain the possible stellar radius and mass values as a function of distance, anchoring model predictions to empirically-observed trends in stellar populations. The brown dwarf mass and radius are in turn constrained by such trends, given the observed radial velocity and transit events. All other base parameters except $T_{\text{eff}}$ are constrained by available measurements from \cite{Johnson_2011} or \cite{Montet_2015}.

\begin{figure*}[!ht]
\centering
\includegraphics[scale=0.43]{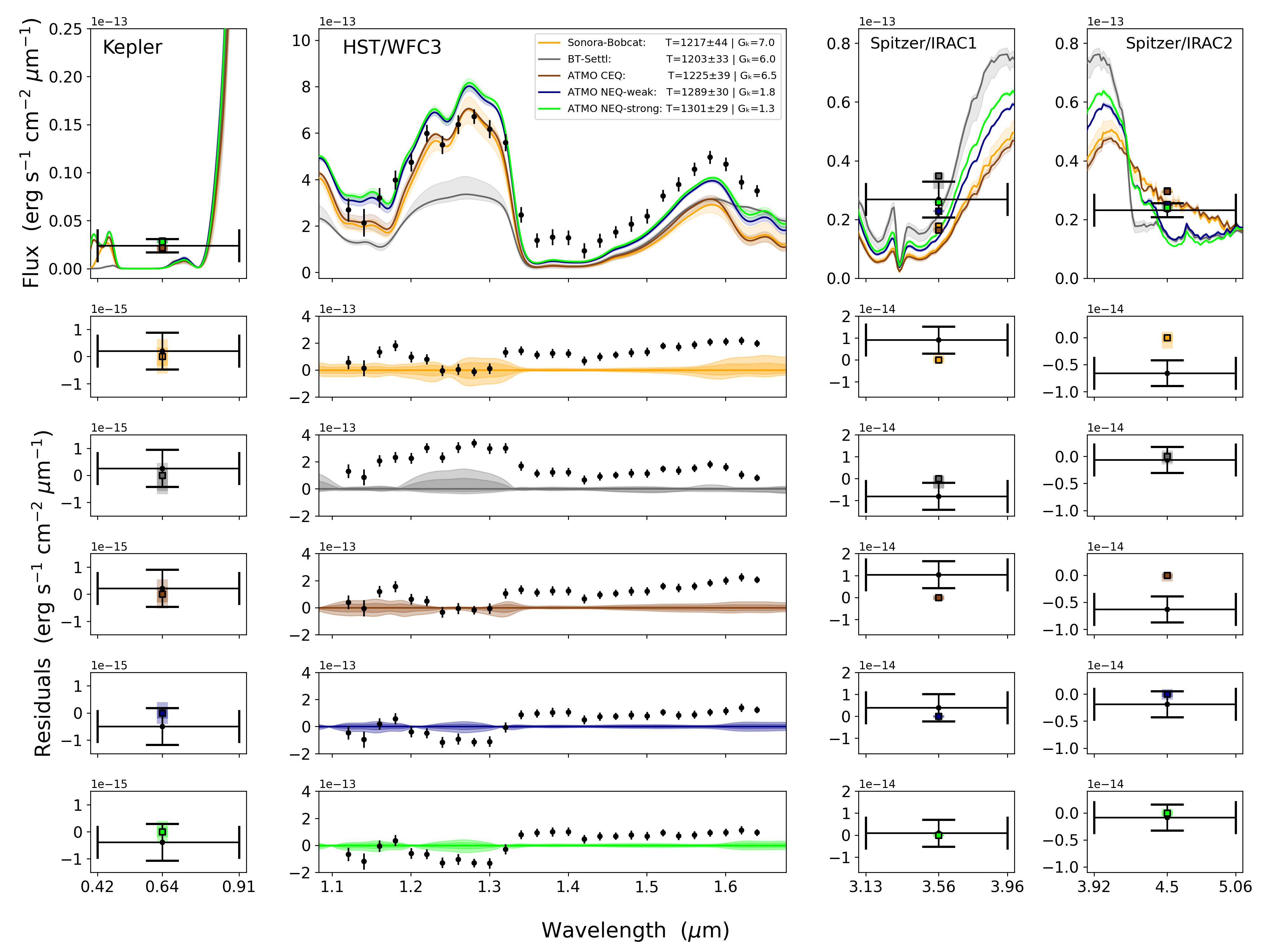}
\caption{ Atmospheric models fit to the full SED of LHS~6343\,C (black) with the surface gravity $\g=GM/R^2$ and scale $(R/d)^2$ constrained to measured values dependent on the Gaia DR3 distance to the system. The second column represents the HST/WFC3 spectrum, while the first, third and fourth columns represent the Kepler and Spitzer/IRAC 1-2 photometry. The orange line represents the Sonora-Bobcat best-fitting median model, the gray one represents the BT-Settl CIFIST11 models, while the brown, blue and green ones represent those for the ATMO-2020 CEQ, NEQ-weak and NEQ-strong models respectively. Additionally, the legend shows the inferred effective temperature distribution and the peak value of the $G_k$ distribution for each model. Each median line is bounded by its $1\sigma$ confidence interval in the first row, while the residuals in subsequent rows also show the $2\sigma$ confidence. Data residuals are shown as an absolute difference relative to models. The model spectra are shown convolved to the resolution of HST/WFC3 across all bandpasses for clarity, even if this convolution is not applied to the photometry prior to binning. The photometric columns are also accompanied by the binned model spectra for each bandpass to facilitate visual comparison with the observed data. Horizontal error bars in the photometric columns represent the wavelength coverage of each data point.\\\text{ }
}
\label{fig:atmos-fit}
\end{figure*}

\subsubsection{\textbf{Atmosphere Modelling With Fixed Distance}}
\label{subsubsec:constrainModFit}

The analyses of this section assume the reported Gaia distance of $\sDIST$ is accurate. As mentioned in section~\ref{subsec:fit-method}, the parameters that characterize an atmospheric model are $T_{\text{eff}}$, the surface gravity, $\g$, and the flux scaling parameter (+ metallicity for Sonora-Bobcat). Surface gravity is solely dependent on the BD's mass and radius ($\g=GM/R^2$), while the scale is solely dependent on the BD's radius and the LHS 6343 system's distance, scaling as $(R/d)^2$. Both the mass and radius are ultimately dependent on the distance and other orbital and photometric parameters described in sections~\ref{subsec:update-stella}~\&~\ref{subsec:fit-method}. We are thus able to constrain everything but $T_{\text{eff}}$ using these measurements. We use a Monte Carlo procedure where, for each step in the MC chain, every base parameter but $T_{\text{eff}}$ is independently pulled from their respective measured normal distributions by \cite{Montet_2015} or \cite{Johnson_2011}. The best-fitting model $k$ is then determined at each step by letting $T_{\text{eff}}$ be the only free parameter in a weighted, non-linear least-squares fit to the spectrum following the weighting scheme of Eq.~\ref{eq:Gk}. Such a framework allows for the models to be truly constrained by the desired parameters, whereas a Bayesian (MCMC) approach would allow for certain parameter chains to possibly deviate from their measured distributions in order to better fit the data.

This approach produces a good overall agreement with the observed emerging flux of the SED for the ATMO-2020 non-equilibrium chemistry (NEQ) models, while the chemical equilibrium (CEQ) models of ATMO-2020 and Sonora-Bobcat and the BT-Settl cloud models reveal more significant tensions in comparison. The results of these constrained atmospheric fits to the entire set of available emission data are shown in Fig.~\ref{fig:atmos-fit}. The best-fit $T_{\text{eff}}$ values and the average $G_k$ statistics for each interpolated model grid are also shown in the legend of Fig.~\ref{fig:atmos-fit}, as well as in Table~\ref{tab:atmos-diagnostics}.

\vspace{10pt}
\noindent
\emph{--- Chemical Non-Equilibrium Model Fits}
\vspace{5pt}

The better fit is provided by the ATMO-2020 NEQ models and occurs at $T_{\text{eff}}\approx 1300\pm30\,\KELVIN$. This is roughly in agreement with that of the spectral type - $T_{\text{eff}}$ empirical relation of \cite{filippazzo_2015} which yielded $T_{\text{eff}} \approx 1192\pm117\,\KELVIN$, about 100\,K more at 1-$\sigma$. For the best-fitting NEQ model, the largest departures observed for photometric observations are smaller than 10-15\%. On the other hand, spectroscopic (i.e. WFC3) residuals can worsen considerably depending on the wavelength range considered. In particular, the $1.4~\mu$m water absorption feature is underpredicted across all models, with even the best-fitting NEQ model being $\sim66\%$ smaller than what is observed by WFC3 in that bandpass. This feature, as well as the $J$-band and $H$-band peaks, tend to be better reproduced by models with higher $T_{\text{eff}}$, as obtained when fitting only the HST/WFC3 spectrum and not fixing the distance (See Fig.~\ref{fig:atmos-fit-4}). Nevertheless, the overall fit of NEQ models indicates that the important physical parameters, such as $T_{\text{eff}}$, the distance and its dependant parameters (i.e., mass, radius) are self-consistent and known with good fidelity.

\vspace{10pt}
\noindent
\emph{--- Chemical Equilibrium Model Fits}
\vspace{5pt}

In contrast, ATMO-2020 and Sonora-Bobcat chemical equilibrium (CEQ) models exhibit greater difficulty in reproducing the totality of the observed SED, resulting in $T_{\text{eff}}\sim1220\pm40\,\KELVIN$. This is unsurprising, as the atmospheric processes of an L/T transition BD such as LHS~6343\,C are not expected to be correctly reproduced by such models. When constrained by the full SED and the prior knowledge on the system distance, Fig.~\textbf{\ref{fig:atmos-fit}} shows they underpredict the Spitzer/IRAC1 and $H$-band regions while overpredicting the Spitzer/IRAC2 data. For the Spitzer discrepancies specifically, similar tension is seen with the pan-chromatic SED fitting of another L/T transition template, HN Peg B \citep{suarez.2021}. CEQ models happen to reproduce the magnitude of the $J$-band peak of the WFC3 spectrum better than NEQ models in this instance, however this is more so a consequence of attempting to optimally fit the heavily weighted Spitzer photometry, which tends to decrease their inferred $T_{\text{eff}}$ and therefore the scaling of the $J$-band feature. Additionally, the inferred $T_{\text{eff}}$ distribution for Sonora-Bobcat has a slight positive correlation with metallicity.

\vspace{10pt}
\noindent
\emph{--- BT-Settl Model Fits}
\vspace{5pt}

The distance-constrained BT-Settl models provide a slightly bi-modal distribution for effective temperature centered at $\sim1200\,\KELVIN$, with peaks at $\sim1170\,\KELVIN$ and $\sim1230\,\KELVIN$. The bi-modal outcome stems from the behaviour of the high surface gravity (e.g., log$\,\g=5.5$) BT-Settl models, which do not necessarily increase in flux as a function of $T_{\text{eff}}$ within the observed Kepler, HST/WFC3 and Spitzer/IRAC 1-2 bandpasses. The BT-Settl models do reproduce the Spitzer photometry better than CEQ models, but still underperform compared to NEQ models. However, the fit to the Spitzer data comes at the expense of adequately reproducing the $J$-band peak of the HST/WFC3 spectrum, where the attenuating effect of clouds on emitted flux that this model implements is not compatible with observations.

\vspace{10pt}
\noindent
\emph{--- Panchromatic Fitting vs. Subset Fitting}
\vspace{5pt}

Fitting to a reduced set of instrument data, namely only the HST/WFC3 spectrum, does allow CEQ and NEQ models to converge to similar values of $T_{\text{eff}}\approx 1330\pm30\,\KELVIN$. However, this causes CEQ models to now overpredict the Kepler and Spitzer/IRAC2 photometry, and the previously mentioned tensions in the WFC3 spectrum (e.g., the $J$-band peak and $1.4~\mu m$ absorption) remain unimproved for all models. The BT-Settl models become a completely separated bimodal distribution for $T_{\text{eff}}$, with narrow spikes in the distribution at 1400 and 1500\,\,\KELVIN. These are both models that are supplied by the base, non-interpolated BT-Settl model grid. The lack of spread in both $T_{\text{eff}}$ distributions again stems from the behaviour of the high surface gravity BT-Settl models, which do not strictly increase in flux between $1200$ to $1600\,\KELVIN$ in the HST/WFC3 bandpass. Therefore, the vast majority of input parameter combinations often converge to the local emission minima or maxima these models offer, as interpolation is anchored to these $T_{\text{eff}}$ grid values.

If all models are instead fit only to the Spitzer data of \cite{Montet_2016}, the interpolated CEQ grids converge roughly to $T_{\text{eff}}\approx1100\pm50\,\KELVIN$. NEQ models still retain roughly the same predictions as in prior cases where all instruments were used, with the inferred $T_{\text{eff}}$ staying within 1-$\sigma$ of the constrained fit with all available data. BT-Settl models predict $T_{\text{eff}}\approx1180\pm25\,\KELVIN$, which remains consistent with the BT-Settl fit using all available data. It also provides a goodness-of-fit $G_k$ similar to NEQ models, illustrating a clear tension between their $T_{\text{eff}}$ predictions obtained using Spitzer.\\

In summary, distance-constrained fits performed with ATMO-2020 NEQ models are the most successful in reproducing the overall SED of LHS~6343\,C, as well as being those with minimal variability between predictions when different instrument subsets are considered. However, no constrained model demonstrates the ability to reproduce all details of the HST/WFC3 spectrum; generally underpredicting the $H$-band spectrum and the $1.4~\mu m$ water absorption feature. Constrained BT-Settl models perform well at reproducing the Spitzer photometry, but fail to reproduce the HST/WFC3 spectrum and more specifically, the general profile of the $J$-band peak.

\begin{deluxetable*}{l|cc|ccccccc}[!ht]

\tabletypesize{\scriptsize}
\tablewidth{0pt} 
\tablecaption{Atmosphere Modelling Results With Constrained and Unconstrained Distances} \label{tab:atmos-diagnostics}
\tablehead{
\colhead{Model \& Data Used }& \colhead{$G_k$} & \colhead{$T_{\text{eff}}$}  & \colhead{$G_k$} & \colhead{$T_{\text{eff}}$} & \colhead{distance} & \colhead{mass} & \colhead{radius} & \colhead{log\,$\g$~ ($\frac{GM}{R^2}$)} & \colhead{scale~ $(\frac{R}{d})^2$}\\[-5pt]
\colhead{} & \colhead{} & \colhead{($\KELVIN$)} & \colhead{} & \colhead{($\KELVIN$)} & \colhead{($\PARSEC$)} & \colhead{($\massJUP$)} & \colhead{($\radiusJUP$)} & \colhead{(cgs)} & \colhead{($\times~10^{-21}$)}
}

\startdata
{}&{}&{}&{}&{}&{}&{}&{}&{}&{}\\
 & \multicolumn{2}{c|}{\textbf{Gaia Distance}} & \multicolumn{7}{c}{\textbf{9~pc~$\leq$~Distance~$\leq$~90~pc~~}} \\[10pt]
\underline{\textbf{Gaia-dependant values}}   &{------}&{------} & {------} & {------} &  {$35.67\pm1.77$}  &  {$62.6\pm2.2$}  &  {$0.788\pm0.043$}  &  {$5.40\pm0.04$}  &  {$2.62\pm0.21$}  \\[15pt]
\underline{\textbf{All Data}}   &{}&{}&{}&{}&{}&{}&{}&{}&{}\\[5pt]
ATMO-2020 NEQ-Strong & $1.3$ & $1301\pm 29$  & $1.1$ & $1346\,^{+52}_{-56}$ & $46.9\,^{+12.9}_{-11.8}$ & $72.5\,^{+8.6}_{-10.7}$ & $0.97\,^{+0.2}_{-0.2}$ & $5.28\,^{+0.13}_{-0.12}$ & $2.28\,^{+0.34}_{-0.29}$ \\[5pt]
ATMO-2020 NEQ-Weak   & $1.8$ & $1289\pm 30$  & $1.4$ & $1355\,^{+50}_{-57}$ & $50.6\,^{+13.0}_{-11.4}$ & $75.1\,^{+7.9}_{-9.1}$ & $1.02\,^{+0.2}_{-0.19}$ & $5.25\,^{+0.12}_{-0.11}$ & $2.19\,^{+0.31}_{-0.28}$\\[5pt]
ATMO-2020 CEQ        & $6.5$ & $1225\pm 39$ & $2.9$ & $1389\,^{+43}_{-49}$ & $68.3\,^{+10.0}_{-11.4}$ & $84.8\,^{+4.2}_{-6.0}$ & $1.25\,^{+0.14}_{-0.16}$ & $5.13\,^{+0.09}_{-0.07}$ & $1.8\,^{+0.2}_{-0.17}$ \\[5pt]
Sonora-Bobcat        & $7.0$ & $1217\pm 44$ & $3.4$ & $1362\,^{+41}_{-47}$ & $62.7\,^{+11.6}_{-11.9}$ & $82.2\,^{+5.4}_{-7.1}$ & $1.18\,^{+0.16}_{-0.18}$ & $5.17\,^{+0.11}_{-0.09}$ & $1.89\,^{+0.23}_{-0.2}$ \\[5pt]
BT-Settl             & $6.0$& $1203\pm 33$ & $5.7$ & $1171\,^{+55}_{-27}$ & $32.1\,^{+6.1}_{-4.6}$ & $59.1\,^{+6.6}_{-5.8}$ & $0.74\,^{+0.12}_{-0.1}$ & $5.42\,^{+0.08}_{-0.08}$ & $2.86\,^{+0.25}_{-0.31}$\\[5pt]
BT-Settl (mode $\#2$) & ------ & ------ & $1.8$ & $931\,^{+10}_{-11}$ & $10.6\,^{+1.1}_{-0.7}$ & $30.4\,^{+1.7}_{-1.0}$ & $0.3\,^{+0.03}_{-0.02}$ & $5.91\,^{+0.04}_{-0.06}$ & $4.39\,^{+0.3}_{-0.29}$ \\[5pt]
\underline{\textbf{HST/WFC3}}&{}&{}&{}&{}&{}&{}&{}&{}&{}\\[5pt]
ATMO-2020 NEQ-Strong & $6.3$ & $1327\pm 29$  & $2.5$ & $1433\,^{+75}_{-86}$ & $63.5\,^{+14.6}_{-18.9}$ & $82.4\,^{+6.3}_{-12.0}$ & $1.19\,^{+0.18}_{-0.27}$ & $5.16\,^{+0.15}_{-0.09}$ & $1.89\,^{+0.45}_{-0.27}$ \\[5pt]
ATMO-2020 NEQ-Weak   & $6.6$ & $1327\pm 29$  & $2.3$ & $1439\,^{+77}_{-81}$ & $64.3\,^{+14.3}_{-18.5}$ & $82.8\,^{+6.1}_{-11.4}$ & $1.19\,^{+0.18}_{-0.26}$ & $5.16\,^{+0.15}_{-0.09}$ & $1.89\,^{+0.41}_{-0.27}$ \\[5pt]
ATMO-2020 CEQ        & $6.8$ & $1335\pm 28$  & $2.1$ & $1446\,^{+73}_{-81}$ & $64.5\,^{+14.2}_{-18.4}$ & $82.9\,^{+6.0}_{-11.1}$ & $1.2\,^{+0.17}_{-0.26}$ & $5.16\,^{+0.15}_{-0.09}$ & $1.88\,^{+0.41}_{-0.26}$ \\[5pt]
Sonora-Bobcat        & $9.9$ & $1324\pm 29$  & $4.8$ & $1465\,^{+86}_{-100}$ & $68.1\,^{+12.3}_{-19.3}$ & $84.7\,^{+4.9}_{-11.0}$ & $1.24\,^{+0.15}_{-0.26}$ & $5.14\,^{+0.14}_{-0.08}$ & $1.8\,^{+0.42}_{-0.24}$ \\[5pt]
BT-Settl             & $10.3$ & $1500\pm 6$ & $7.2$ & $1370\,^{+61}_{-124}$ &  $19.2\,^{+11.2}_{-6.6}$ &  $42.5\,^{+14.7}_{-9.5}$ &  $0.48\,^{+0.23}_{-0.15}$ &  $5.66\,^{+0.21}_{-0.21}$ &  $3.3\,^{+0.47}_{-0.46}$ \\[5pt]
BT-Settl (mode $\#2$)  & $10.8$ & $1400\pm 2$ &------&------&------&------&------&------&------ \\[5pt]
\underline{\textbf{Spitzer/IRAC 1-2}}&{}&{}&{}&{}&{}&{}&{}&{}&{}\\[5pt]
ATMO-2020 NEQ-Strong & $0.02$ & $1291\pm 33$     & $0.03$ & $1336\,^{+100}_{-124}$ & $44.3\,^{+20.0}_{-17.5}$ & $70.4\,^{+12.7}_{-18.0}$ & $0.92\,^{+0.28}_{-0.31}$ & $5.31\,^{+0.22}_{-0.16}$ & $2.35\,^{+0.59}_{-0.47}$\\[5pt]
ATMO-2020 NEQ-Weak   & $0.4$ & $1269\pm 35$        & $0.04$ & $1376\,^{+97}_{-135}$ & $53.3\,^{+19.5}_{-19.4}$ & $76.9\,^{+10.1}_{-16.2}$ & $1.05\,^{+0.26}_{-0.3}$ & $5.23\,^{+0.19}_{-0.14}$ & $2.11\,^{+0.57}_{-0.38}$ \\[5pt]
ATMO-2020 CEQ        & $2.7$ & $1114\,^{+46}_{-41}$ & $0.5$ & $1391\,^{+95}_{-174}$ & $71.0\,^{+10.5}_{-22.0}$ & $85.7\,^{+4.2}_{-11.7}$ & $1.27\,^{+0.12}_{-0.29}$ & $5.12\,^{+0.16}_{-0.07}$ & $1.75\,^{+0.45}_{-0.21}$ \\[5pt]
Sonora-Bobcat        & $2.6$ & $1098\,^{+61}_{-49}$& $0.08$ & $1370\,^{+93}_{-159}$ & $68.7\,^{+11.7}_{-20.2}$ & $84.9\,^{+4.8}_{-11.4}$ & $1.24\,^{+0.14}_{-0.27}$ & $5.14\,^{+0.15}_{-0.07}$ & $1.78\,^{+0.45}_{-0.22}$  \\[5pt]
BT-Settl             & $0.2$ & $1180\pm 25$        & $0.02$ & $1150\,^{+85}_{-104}$ & $34.7\,^{+25.9}_{-11.4}$ & $61.8\,^{+19.5}_{-14.0}$ & $0.78\,^{+0.4}_{-0.22}$ & $5.4\,^{+0.18}_{-0.24}$ & $2.64\,^{+0.48}_{-0.69}$\\[15pt]
\enddata

\tablecomments{The reported $G_k$ represent the peak of the distribution of such values calculated for all interpolated atmospheric models created at each step in a parameter chain. To recover the peak, the $G_k$ distributions are binned to form a histogram and fitted with a skew-normal distribution. Physical parameters calculated using the adopted Gaia DR3 distance are shown in the first row for comparison to those obtained with a uniform prior for distance. For Sonora-Bobcat models explored using an MCMC procedure (section~\ref{subsubsec:unconstrainModFit}), metallicity is given a gaussian prior of $0.03\pm0.26$ informed from the value reported in \cite{Montet_2015}. The resulting distributions for metallicity end up at values of $0.17\pm 0.19$, $0.01\pm 0.21$ and $0.08\pm 0.22$, when fitting all data, HST only and Spitzer only, respectively. All other models have fixed solar metallicity.}
\end{deluxetable*}

\begin{figure*}
    \centering
    \includegraphics[width=0.96\linewidth]{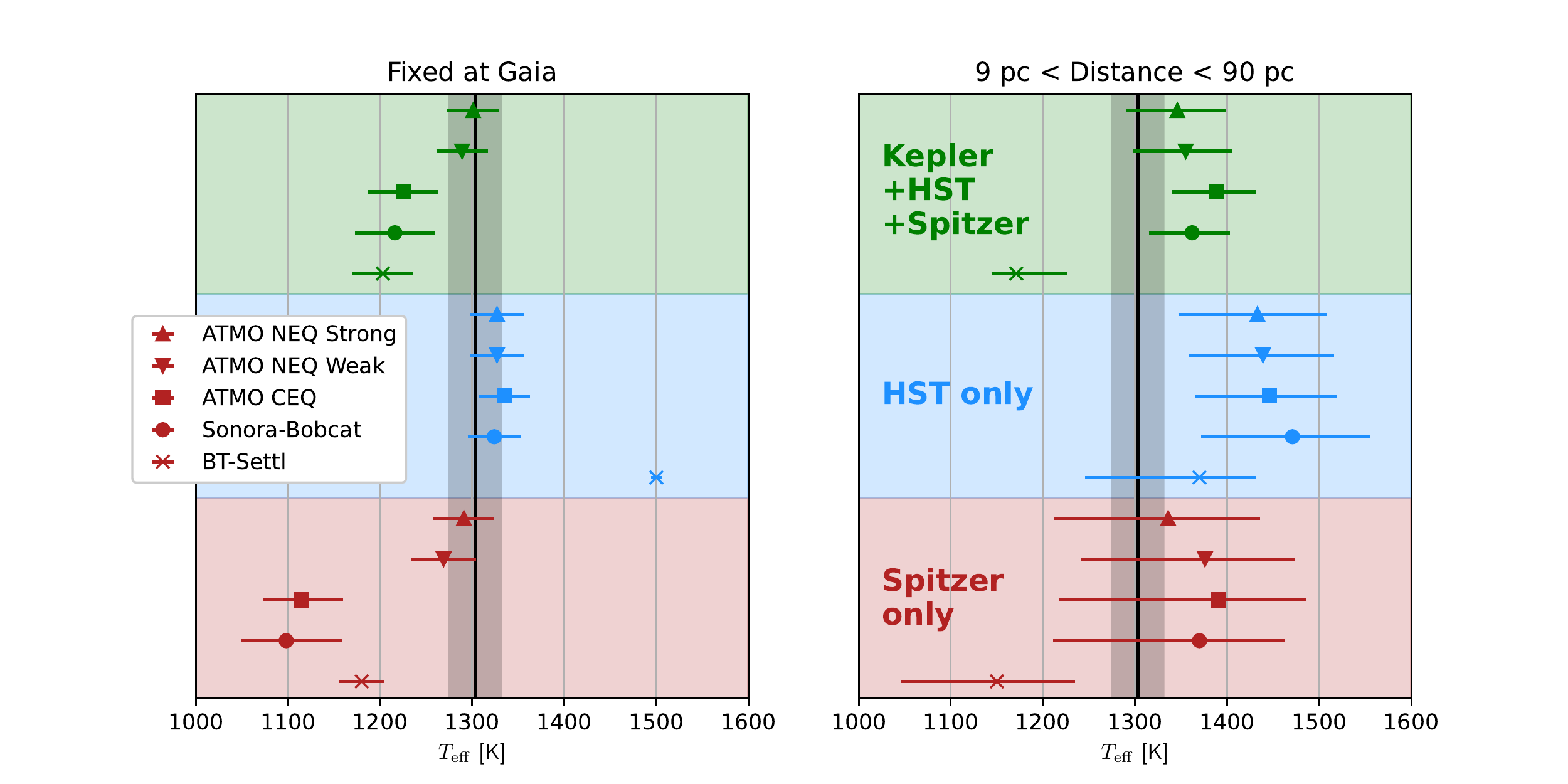}
    \caption{Effective temperature of the best-fit models in the case where the distance was fixed at the Gaia DR3 value, $\sDIST$ (left), and in the case where it was essentially unconstrained with 9~pc~$\leq$~distance~$\leq 90$~pc (right). For each of the 5 models used, the fit is either performed using all the data sets, i.e. secondary eclipses with Kepler, HST and Spitzer (top - green), using only the HST eclipse (middle - blue) or only the Spitzer eclipses (bottom - red). Horizontal error bars represent 1-$\sigma$ confidence intervals. The grey vertical band represents the temperature obtained when applying the Stefan-Boltzmann equation from the measured SED, assuming the Gaia distance and associated BD radius.}
    \label{fig:fig_teffcomparison}
\end{figure*}

\vspace{-0.8cm}
\subsubsection{\textbf{Atmosphere Modelling With Free Distance}}
\label{subsubsec:unconstrainModFit}

Given that all distance-constrained fits systematically show some level of disagreement within the HST/WFC3 wavelength range, and that the reported Gaia DR3 distance may be misrepresented, we explore leaving the system distance as a free parameter during the fit. Naturally, doing so ignores the empirical, independently measured prior on this parameter, but it allows insight into what the models predict given fewer constraints on the fit; and if those predictions match with Gaia-dependent measurements. Having a free scale parameter, $(R/d)^2$, is also standard procedure when fitting the spectra of field BDs which usually don't provide a way to directly measure their radius. If the system distance reported by Gaia is assumed correct, then the following unconstrained analyses can be regarded as a test of brown dwarf atmospheric models, where departures from the constrained fits of section~\ref{subsubsec:constrainModFit} would mean that model predictions would have been different had LHS 6343 C been observed as a field BD. Otherwise, the inferred distances, and subsequent radii and masses, of the unconstrained fits can reveal model-dependant bounds for the distance to LHS~6343.

To implement unconstrained model fits, we adopt an MCMC procedure exploring the parameter space detailed in section~\ref{subsec:fit-method}. $T_{\text{eff}}$ and the distance are given independent uniform priors, which for the latter implies that the brown dwarf mass, radius, surface gravity and flux scale also have uniform, albeit correlated, priors. The uniform prior on the distance is informed from the possible absolute magnitudes that can be input to the \cite{Mann_2015} and \cite{Mann_2019} stellar empirical relations for M dwarfs, which is assumed to be $M_{K_S}=[4.6\, ,\,9.3]$ within this work. This effectively results in a distance prior of roughly $[9\, ,\, 90]~\PARSEC$. The orbital period of LHS~6343\,C around its host star is fixed, as its uncertainty is negligible. The remaining free parameters are given independent Gaussian priors about their reported values by \cite{Montet_2015} or \cite{Johnson_2011}. Finally, the log-likelihood for this MCMC is described by a variant of Eq.~\ref{eq-likemike} which incorporates the goodness-of-fit principle of Eq.~\ref{eq:Gk}, and convergence is evaluated the same way as described in section~\ref{subsubsec:mcmc-mod}. The MCMC fit diagnostics and inferred physical parameters for every model and instrument dataset studied are found in Table~\ref{tab:atmos-diagnostics}.

Fig.~\ref{fig:atmos-fit-3} shows the results of such MCMC atmospheric model fits applied to the full LHS~6343\,C spectrum (Kepler, HST, Spitzer). This approach yielded $T_{\text{eff}}\approx 1380\pm 50\,\KELVIN$ for the CEQ models and $T_{\text{eff}}\approx 1350\pm 54\,\KELVIN$ for NEQ models. The increased $T_{\text{eff}}$ of CEQ models is made possible by a notable increase in distance, mass and radius, allowing them to output a fit more closely aligned with the observed Spitzer photometry and the $H$-band peak of the HST/WFC3 spectrum. The inferred $[Fe/H]$ metallicity of Sonora-Bobcat models increased to $0.17\pm0.19$ compared to the Gaussian prior of $0.03\pm0.26$ informed from the value reported by \cite{Montet_2015}. In addition, the positive correlation of metallicity with $T_{\text{eff}}$ is no longer observed. However, the associated physical parameters become unrealistic. The inferred masses venture close or above the substellar threshold (70 to 85\,$\massJUP$) while radii are $\approx$\,50\% higher than previously estimated for LHS 6343 C, as they imply either low-mass main sequence stars or young ($<1\,Gyr$) BDs respectively.

The slight bi-modal distribution in $T_{\text{eff}}$ observed for BT-Settl models in section~\ref{subsubsec:constrainModFit} now becomes a fully separated bi-modal distribution with peaks at $1171\,^{+55}_{-27}$ and $931\pm11\,\,\KELVIN$. However, the second peak is at such low $T_{\text{eff}}$ and distance (and therefore mass and radius) that it is irreconcilable with known characteristics of LHS~6343\,C. The radius is especially irreconcilable with any BD mass according to the evolutionary models used in this work.

Therefore, the atmospheric model with the best goodness-of-fit $G_k$ remains the ATMO-2020 NEQ-strong grid in this instance. It is also one of two models with physical parameter predictions (e.g., mass, radius) that are consistent within 1-$\sigma$ of those calculated using the reported Gaia DR3 distance. The CEQ models of both ATMO-2020 and Sonora-Bobcat predict larger distances, and thus larger masses and radii which are not representative of a T1.5 dwarf like LHS~6343\,C. The physical parameters predicted by the $1171\,^{+55}_{-27}\,\,\KELVIN$ BT-Settl model do technically align more with Gaia-based measurements, but the $G_k$ of this particular fit is significantly worse ($\sim6$) compared to that of the ATMO-2020 NEQ-strong value ($\sim1$). 
Finally, the fit to the 1.4~$\mu$m water band is slightly improved across all models. However, it remains underpredicted, with emission values being at least $\sim45\%$ smaller than what is observed.

\begin{figure*}[!t]
\centering
\includegraphics[scale=0.43]{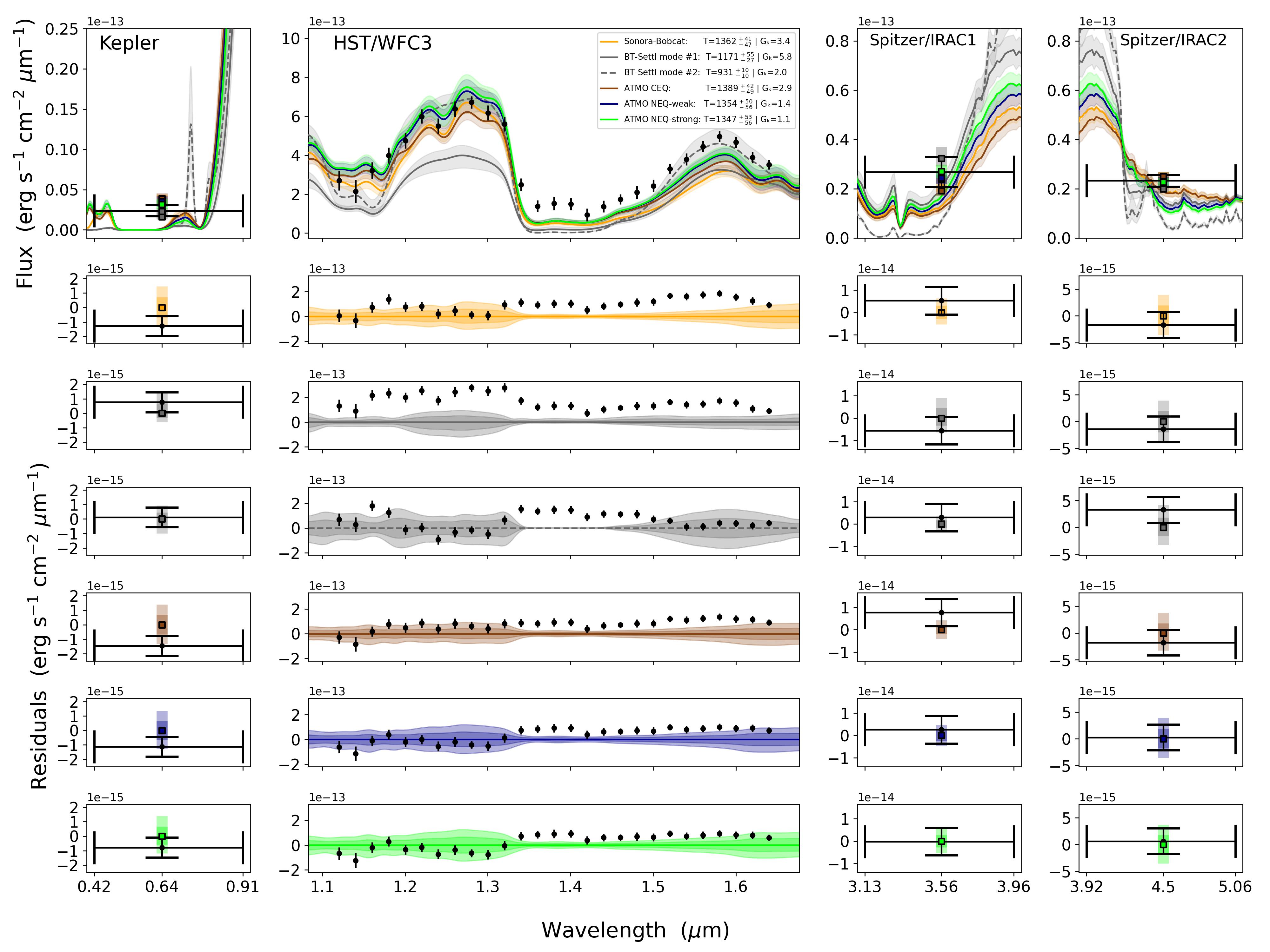}
\caption{Atmospheric models fit to the full SED of LHS~6343\,C with the system distance prescribed a uniform prior (9~pc~$\leq$~distance~$\leq$~90~pc). Refer to the caption of Fig.~\ref{fig:atmos-fit} for details on the plot layout.\\\text{}}
\label{fig:atmos-fit-3}
\end{figure*}

Fitting the unconstrained atmospheric models to only the HST/WFC3 spectrum has the effect of significantly increasing the $T_{\text{eff}}$ of all models. In fact, the upper limit of the uniform prior on the distance significantly caps the $T_{\text{eff}}$ predictions of these models. All ATMO-2020 and Sonora-Bobcat models saw their predictions rise to $\sim1450\pm80\,\KELVIN$, representing an increase of  $\sim100\,\KELVIN$ compared to unconstrained fits performed on all available emission data. BT-Settl models reach $T_{\text{eff}}=1363\,^{+61}_{-128}\,\KELVIN$. These fits do allow for a more robust fitting of the 1.4~$\mu$m water band and the $J$ and $H$-band peaks. However, none of the inferred distances are in agreement with the reported Gaia distance, hovering around $65\pm16\,\PARSEC$ for CEQ and NEQ models and $18\pm10\,\PARSEC$ for BT-Settl. The inferred masses and radii of these models are in the regime of low-mass main sequence stars and young ($<1\,Gyr$) BDs respectively. In BT-Settl's case, the inferred radius is $\sim0.46\,^{+0.24}_{-0.15}\,R_{\text{Jup}}$, which is irreconcilable with radii predicted at any mass by the evolutionary models used in this work. Therefore, the underlying physical parameters for all unconstrained models fit to only HST/WFC3 can be considered unfeasible.

Fitting the unconstrained atmospheric models to only the Spitzer photometry paints a similar picture for CEQ models, where they require similar higher distances, masses and radii than NEQ models to attain roughly the same $T_{\text{eff}}$. Only the ATMO-2020 NEQ and BT-Settl models provide physical parameters not only consistent with Gaia-dependant measurements but also with evolutionary models as well. However, a similar tension observed in section~\ref{subsubsec:constrainModFit} is apparent here, where the inferred $T_{\text{eff}}$ of these models have roughly a $200\,\KELVIN$ difference while offering a similar quality of fit to the Spitzer photometry.\\

To summarize, when the distance is essentially unconstrained (uniform prior of 9~pc~$\leq$~distance~$\leq$~90~pc), the ATMO-2020 models with non-equilibrium (NEQ) chemistry still offer the best fit. When applied to either the entire set of observations or only to Spitzer photometry, both the unconstrained ATMO NEQ models and the BT-Settl cloudy models infer physical parameters like distance, mass and radius that are consistent within 1-$\sigma$ of values calculated using a Gaia DR3 parallax for the LHS 6343 system. However, their $T_{\text{eff}}$ predictions differ by $\sim200\,\KELVIN$, and the BT-Settl models perform the worst when fit to the entire set of observations. Unconstrained ATMO-2020 and Sonora-Bobcat chemical equilibrium (CEQ) models generally require higher distances, masses and radii that reach the stellar regime to optimally fit observations. Finally, the features of the HST/WFC3 spectrum, like the 1.4~$\mu$m water band, are only adequately fit by models with higher $T_{\text{eff}}$. None of the fits by such models yield physical parameters consistent with Gaia-dependent values and instead yield unfeasible values given the $\sSPTc$ spectral type of LHS~6343\,C.

In fact, the tension between the observed shallow water absorption band at $\geq 1.35~\mu$m and models fit to all available data is the most puzzling. It could perhaps be lifted assuming that opacities at those wavelengths are overestimated and that the emerging flux comes from slightly hotter and deeper atmospheric layers. Alternatively, since water is the dominant opacity at $1.4~\mu$m, it could be that water is less abundant than expected. But, any explanation needs to simultaneously explain why the $H$-band peak, less affected by water opacities, is stronger than all best-fit models.

\begin{figure*}[!ht]
\centering
\includegraphics[scale=0.43]{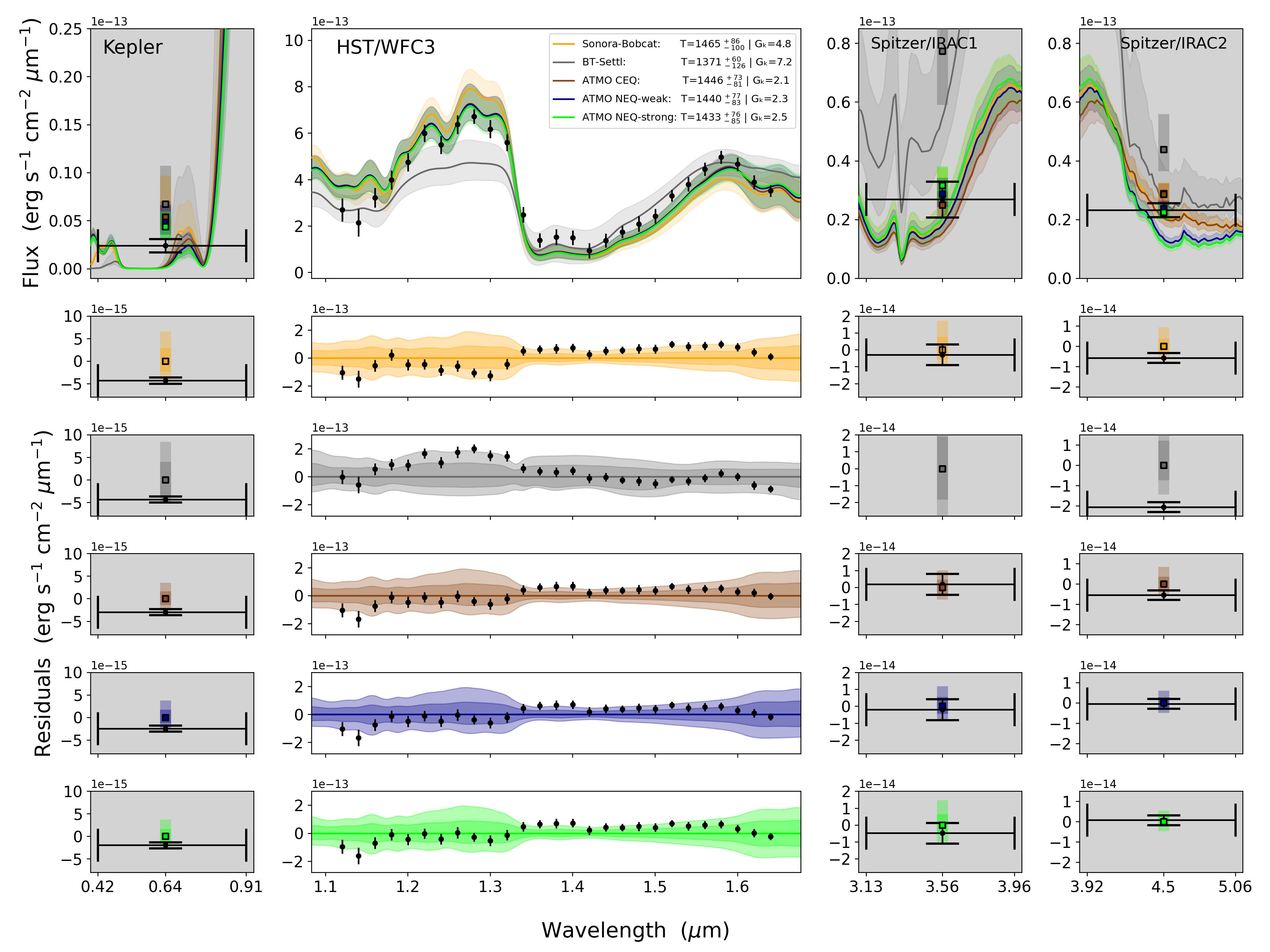}
\caption{Atmospheric models fit only to the HST/WFC3 spectrum of LHS~6343\,C with the system distance prescribed a uniform prior (9~pc~$\leq$~distance~$\leq$~90~pc). Grayed-out columns are used to indicate that the Kepler and Spitzer photometry were not included in the fit. As such, we can still observe what the models predict in those bandpasses. Refer to the caption of Fig.~\ref{fig:atmos-fit} for further details on the plot layout.\\ \vspace{0.2cm}
}
\label{fig:atmos-fit-4}
\end{figure*}

\subsubsection{\textbf{Luminosity and Effective Temperature of LHS~6343\,C}
\label{subsubsec:direct-meas}}

Obtaining direct measurements of LHS~6343\,C's bolometric luminosity, $L_{\text{bol}}$, and effective temperature, $T_{\text{eff}}$, is achieved through the integration of the BD's SED. Since the Kepler, HST and Spitzer bandpasses are not sufficient to fully cover the BD's SED, it becomes necessary to interpolate the gaps using the best-fitting models obtained from section~\ref{subsubsec:unconstrainModFit}. The model used for SED completion is the best-fit ATMO-2020 NEQ-strong model where $T_{\text{eff}}$ and the system distance were given uniform priors in an MCMC analysis. The inferred $T_{\text{eff}}$ for this NEQ-strong fit remains within error of its distance-constrained counterpart, supporting the claim that the chosen model used for the SED interpolation has very little impact on the recovered $L_{\text{bol}}$.

Integrating all observed and modelled apparent fluxes in our chosen SED bandpass (0.42 to 30 $\mu m$) and multiplying the result by a scale factor $(d/R)^2$ (informed from the Gaia distance $d$) yields a total radiant emittance of $j^* = \sBBRc$. Error propagation in the modelled gap regions is performed by taking the mean and standard deviation of gap integrals for each interpolated model in the MCMC chain. The $j^*$ estimate is model-interpolated at the level of $43\pm5$~\% and observation-based at $57\pm6$~\%. Each data set contributes at the following levels: HST ($45\pm5$~\%), Spitzer ($11.4\pm2$~\%) and Kepler ($0.27\pm0.08$~\%). 

Solving the Stefan-Boltzmann equation $j^*=\sigma_{SB} T^4$ yields an effective temperature for the BD of $T_{\text{eff}}=\sTEFFc$. To obtain a bolometric luminosity, the SED integral is instead simply multiplied by the area $4\pi d^2$ which yields log($L_{\text{bol}}/L_{\sun}$) = $\vLBOLc \pm \eLBOLc$, of which models account for 48\% of $L_{\text{bol}}$'s total variance. These new values contrast with those previously reported by \cite{Montet_2016} of $1130\pm50\,K$  and $-5.16\pm0.04$ log($L_{\odot}$) using Spitzer photometry only. This is mostly due to the different $T_{\text{eff}}$ obtained, which by itself represents a luminosity increase of log$((1300/1130)^4)\approx0.24$ dex. Unsurprisingly, the recovered temperature is in close agreement with the NEQ atmosphere model best fit with constrained distance ($T_{\text{eff}} \approx 1300\pm30$) because distance is the driving parameter for that fit.

\subsection{\textbf{The Age of LHS~6343\,C}
\label{subsec:evol}}

The sole physical parameter of LHS~6343\,C for which we lack a direct measurement is age. Assuming that model predictions and the Gaia DR3 system distance are correct, it can be estimated by interpolating evolutionary grids using the BD's measured mass and luminosity as inputs. Interpolation of the grid is performed using the evolutionary routines of the \texttt{SPLAT} package of \cite{burgasser2017spex}. We account for uncertainty in the inputs using a Monte Carlo process, drawing the mass and luminosity values from their respective normal distributions reported in Table~\ref{tab:TheBigOne}. We find the age of LHS~6343\,C to be $\sAGEc$ and $\sAGEcSB$ for the ATMO-2020 and Sonora-Bobcat solar metallicity grids respectively. Naturally, these age estimates are dependent on the Gaia distance assumed when calculating masses with empirical stellar relations. Table~\ref{tab:evol-results} compares the other model predictions (radius, $T_{\text{eff}}$ and $\log\g$) for these 2 model grids to the directly measured values (of Table~\ref{tab:TheBigOne}).

Both model grids can reproduce within 1-$\sigma$ the observed physical properties of LHS~6343\,C (radius, $T_{\text{eff}}$ and log$(\g)$) given its measured mass and luminosity as inputs. The ATMO-2020 grid is slightly better aligned with the observed properties compared to Sonora-Bobcat. One thing to note is the tight distribution of the inferred radius from evolutionary models compared to the measured value from Kepler transit photometry and empirical relations. Table~\ref{tab:updated-pars-error-breakdown} shows that the major contributor to our radius measurement's uncertainty comes from the adopted value for the system distance. The empirical stellar radius-luminosity calibration of \cite{Mann_2015} is also significant in its influence. Thus, better constraints on such empirical stellar relations or the distance would allow for a better test of radius and log$(\g)$ predictions from these evolutionary models.

The sub-solar ($[{\rm Fe/H}]=-0.5$) and super-solar ($[{\rm Fe/H}]=0.5$) metallicity models of Sonora Bobcat do predict lower ($2.70^{+0.37}_{-0.30}$) and higher ($4.23^{+1.33}_{-0.74}$) Gyr ages respectively compared to the solar metallicity models. Even though LHS~6343\,C is measured to have a metallicity close to solar  ($[{\rm Fe/H}]=0.03\pm0.26$), the uncertainty on that measurement does not exclude the possibility of the non-solar values. Thus, such Sonora Bobcat model predictions offer somewhat exaggerated boundaries on the model-dependant age of LHS~6343\,C.

\begin{deluxetable}{lcc}[!t]
\tablewidth{0pt} 
\tablecaption{ LHS~6343\,C Evolutionary Modelling Results \& Comparisons  \label{tab:evol-results}}
\tablehead{ \colhead{Parameter} & \colhead{Predicted Value} & \colhead{Measured Value} }
\startdata
{}&{}&{}\\
\underline{ATMO-2020} & {} & {} \\
Radius [$\radiusJUP$] & $0.805\pm 0.008$ & $\vRADc\pm\eRADc$ \\
$T_{\text{eff}}$ [$\KELVIN$] & $1291\pm 20$ & $\vTEFFc\pm\eTEFFc$ \\
log(\g) [cgs] & $5.379\pm 0.024$ & $\vLOGGc\pm\eLOGGc$ \\
Age [Gyr] & $\vAGEc\eAGEc$ & \\
{}&{}&{}\\
\underline{Sonora-Bobcat} & {} & {} \\
Radius [$\radiusJUP$] & $0.817\pm 0.009$ & $\vRADc\pm\eRADc$ \\
$T_{\text{eff}}$ [$\KELVIN$] & $1280\pm 20$ & $\vTEFFc\pm\eTEFFc$ \\
log(\g) [cgs] & $5.366\pm 0.024$ & $\vLOGGc\pm\eLOGGc$ \\
Age [Gyr] & $\vAGEc\eAGEc$ & \\
{}&{}&{}\\
\enddata
\tablecomments{The "Measured Value" column has the same entries for both model grids, as they are independent of models. They are however influenced by the Gaia DR3 parallax measurement.}
\end{deluxetable}

\begin{deluxetable}{lrrc}[!t]
\tablewidth{0pt} 
\tablecaption{ LHS~6343 Physical Parameters 
\label{tab:TheBigOne}}
\tablehead{
    \colhead{Parameter} & \colhead{Value} & \colhead{1$\sigma$ uncertainty} & \colhead{Reference}
}
\startdata
{}&{}&{}&{}\\
\underline{Stellar Parameters} & {} & {} & {} \\
\nKp              &   \vKp   &   \eKp   &   A \\
\nKpBA            &   \vKpBA   &   \eKpBA   &   B \\
2MASS \nKs              &   \vKs   &   \eKs   &   A \\
2MASS \nKsBA            &   \vKsBA   &   \eKsBA   &   A \\
$M_A$ $[\uMASSa]$       &   \vMASSa   &   \eMASSa   &   ACE \\
$M_B$ $[\uMASSb]$       &  \vMASSb   &   \eMASSb   &   ACE \\
$R_A$ $[\uRADa]$       &   \vRADa   &   \eRADa   &   ADE \\
$R_B$ $[\uRADb]$       &   \vRADb   &   \eRADb   &   ADE \\
$\nTEFFa$ $[K]$       &   \vTEFFa   &   \eTEFFa   &  ADE \\
$\nTEFFb$ $[K]$       &   \vTEFFb   &   \eTEFFb   &   ADE \\
System Distance $[\uDIST]$      &   \vDIST   &   \eDIST   &   E \\
System Metallicity [Fe/H]   &   0.03   &   0.26   &   B \\
System Metal Content [a/H]  &   0.02   &   0.19   &   B \\[5pt]
\underline{Brown Dwarf Parameters} & {} & {} & {} \\
MKO $M_J$                      &   \vKsC  &   \eKsC  &  \\
MKO $J-H$                      &  0.76    &   0.06  & F \\
$M_C$ $[\uMASSc]$                &   \vMASSc   &   \eMASSc   &   ABCE \\
$R_C$ $[\uRADc]$                &   \vRADc   &   \eRADc   &   ABDE \\
Sp. Type (NIR)                   &   \vSPTc   &   \eSPTc     &   G \\
$\nTEFFc$ $[K]$         &   \vTEFFc   &   \eTEFFc   &   H \\
$L_{\text{bol}}$ $[\uLBOLc]$   &   \vLBOLc   &   \eLBOLc   &   H \\
Age $[\uAGEc]$                      &   \vAGEc   &   $\eAGEc$   &   I \\
Mean Density $\rho_C$ $[\uRHOc]$  &  \vRHOc  &  \eRHOc  &  ABCDE \\ 
Surface Gravity log$(\g_C)$ [cgs]  &  \vLOGGc  &  \eLOGGc  &  ABCDE \\ 
Semimajor Axis $[\uSMAc]$  &  \vSMAc  &  \eSMAc  &  ABDE \\
$T_{eq}$\text{\text{ } } $($ $T_{\text{eff},A}$ $(\frac{R_A}{2a})^{\frac{1}{2}}$ $)$ $[K]$  &  \vTEQc  &  \eTEQc  &  ABDE \\
{}&{}&{}&{}\\
\enddata

\tablecomments{\text{ }References as to the contributors to each listed value are described below: \vspace{0.1cm}\\
A) From \cite{Johnson_2011}'s discovery paper, Table 1.\\
B) From \cite{Montet_2015} Tables 2, 3.\\
C) \cite{Mann_2019}'s Mass-Luminosity Relation.\\
D) \cite{Mann_2015}'s Radius-Luminosity, Radius-$T_{\text{eff}}$ relations.\\
E) Gaia DR3 parallax for LHS~6343\,A with inflated uncertainty.\\
F) From \cite{Manjavacas_2013}, assuming the $J-H$ color of 2MASS J11061197+2754215 as an analog to that of LHS~6343\,C\\
G) Spectral typing using the \texttt{SPLAT} python package.\\
H) Semi-empirical measure using the ATMO-2020 NEQ-strong atmospheric model grid, as well as contributors ABCDE.\\
I) Model-dependant measure using the ATMO-2020 NEQ-strong evolutionary model grid, as well as contributors ABCDE.
}
\end{deluxetable}

\section{\textbf{Discussion}}

\subsection{\textbf{Exploring Alternate Distances to LHS~6343}}
\label{subsec:alt-d}

Until a more robust parallax fit to the LHS~6343 stellar binary is released by the Gaia consortium, the distance to LHS~6343~C remains uncertain. In the meantime, we explore independent distance estimates in this section.

The first avenue relies on adopting stellar evolutionary models. \cite{Montet_2015} have used Dartmouth stellar models within their MCMC framework to infer a distance of $32.7\pm 1.3\,\PARSEC$ given the resolved photometry on both stellar components. \cite{Johnson_2011} arrived to an estimate of $36.6\pm1.1$~pc using the Padova atmosphere models, a difference of about 2 sigmas from each other and roughly consistent with the tentative Gaia DR3 measurement of $\sDIST$. Our study attempts to infer physical parameters with as few model assumptions as possible so we did not explore this option.

The second option is to use spectral type - magnitude empirical relations in the stellar regime. However, no reliable measurement of spectral type for any of the LHS~6343 M dwarfs has seemingly been published: \cite{Reid_2004} have M2.5 for the primary M dwarf, \cite{Fouque_2017} state M3.6+M3.6, \cite{Herrero_2013} adopt M4+M5. The first estimate seems to have used the combined light of A \& B for spectral typing. The second estimate yields equal spectral types for A and B, which is unlikely based on their $\approx 0.5$ magnitude difference in the NIR bands and their $\sim100\,K$ temperature differences reported by \cite{Montet_2015} and this work. Additionally, spectral type - magnitude empirical relations are very steep in the early M dwarf range, which widens the recovered magnitude range.

For the sake of recovering a photometric distance for component A, a spectral type of M$3.6\pm0.5$ is assumed. A high-order polynomial fit to the $M_G$ vs. spectral type sequence is built from the Gaia DR3 photometry of a sample of field stars with high-quality parallaxes (J. Gagné et al., in preparation), based on the same methodology as the spectral type - color relation described in \cite{Gagn__2020}. The sample includes stars within 100 pc having a spectral type listed in Simbad and not belonging to any of the kown young associations. The scatter of the Gaia $M_G$ vs. spectral type relation is 0.71 mag. Since metallicity is responsible for some of that spread, the relation is further corrected by selecting 80 stars with accurate metallicity from \cite{Mann_2015}, spanning the M0 to M4 spectral types and with the same metallicity as LHS~6343\,A (Z=+0.03$\pm$0.26). That sample has no offset with respect to the Gaia relation but a smaller scatter of 0.33 mag. In the end, this method yields a distance for LHS~6343\,A of $27.2\,^{+9.3}_{-6.9}\,\PARSEC$. This value remains 1$\sigma$-consistent with the Gaia DR3 value, but has a probability distribution with a noticeably smaller peak value and larger uncertainties. Turning the problem around, we note that, given the Gaia distance, the $M_G$ vs. spectral type relation retrieves a spectral type of M3.1$\,^{+0.6}_{-0.7}$, consistent with our adopted estimate of M3.6$\pm0.5$. In other words, there is no significant tension between distance, magnitude and spectral type with this method. 


A third possibility is using the L/T transition itself as a standard candle and placing LHS~6343~C at the median of that distribution. In fact, LHS~6343~C appears underluminous in $M_J$ compared to the similar spectral type BDs found in Fig. \ref{fig:CMD}, which could be the result of having underestimated the distance rather than being an intrinsic property. We obtain a photometric distance based on a polynomial fit to the $M_J$ vs. spectral type data found within the UltracoolSheet of \cite{best_2020}. The parameter space of this polynomial fit is explored using an MCMC technique where the parameters are allowed to vary along the sequence, which tends to reproduce well the unusual shape of the CMD at those spectral types (J. Gagné et al., in prep). Using the input spectral type of $\sSPTc$, the resulting absolute $J$ magnitude becomes $M_J\approx14.1\pm0.4$. Combined with the measured $J$-band apparent magnitude from the HST/WFC3 spectrum ($17.632\pm0.025$), it implies a photometric distance of $50.1\,^{+10}_{-8.6}\,\PARSEC$, which is more than 1-$\sigma$ away from the Gaia DR3 distance of $\sDIST$. We caution that this magnitude - spectral type relation for L/T brown dwarfs should not be regarded to be as reliable as the same relation in the stellar regime because it is based on a heterogeneous sample of objects: since no main sequence exist for BDs, objects observed at the L/T transition span a wide range of masses and ages.

Interestingly, the best fits of our own SED fitting (NEQ models with the distance left free) favor larger distances of $\approx 40-50~pc$, but still technically remain within 1-$\sigma$ of the Gaia value. Distances higher than this range lead to unlikely solutions according to empirical stellar relations and evolutionary models; e.g. masses above the substellar limit or radii up to 30-50\% larger than estimated based on transit depths. Thus, distances for LHS~6343\,C based on the L/T transition spectral type - magnitude relations could be overestimated, while analogous relations for M dwarfs do not lead to those pitfalls.

Note that the increased distances obtained from substellar atmospheric modelling are more consistent with the photometric distance obtained from an L/T spectral type - magnitude empirical relation ($\sim50\pm10\,\PARSEC$), and much less so with a photometric distance obtained using a similar relation in the M dwarf regime ($\sim27\pm8\,\PARSEC$).


\subsection{\textbf{Modelling the L/T transition}}

With a $\sSPTc$ spectral type, LHS~6343\,C is within the L/T transition for BDs. This transition is observed in colour-magnitude diagrams as an increased $J$-band magnitude both absolutely and relative to $H$ or $K$ bands (i.e., bluer colours) for early T dwarfs. Current hypotheses as to the nature of this transition point to two possible mechanisms. One of them is the presence of clouds as having a significant impact on the observed properties of L/T transition BDs \citep{Allard_2001}. Dust clouds are thought to be a major source of opacity in the near-IR for L dwarfs. Their temperature and pressure favour the appearance of magnesium silicate and iron condensates, causing them to redden as this dust further accumulates in later spectral types. However, this trend is temporarily reversed once temperatures drop below $\sim 1400$K; the near-IR colours of early and mid-T dwarfs become noticeably bluer compared to late-L types. Cloud models suggest this transition is due to the disappearance of clouds below the observable photosphere \citep{allard2013progress}. A physical explanation is presented by \cite{Tan_2019}, where the authors show that clouds at the L/T transition with larger particles dissipate more easily than those with smaller ones. Previous modelling efforts seemed to suggest that L dwarfs are dominated by sub-micron particles, while T dwarfs would possess larger particle sizes \citep{Saumon_2008,Burningham_2017}. The dissipation of a photospheric cloud deck would then be a natural occurrence for BDs transitioning from L to T types.

The other proposed mechanism to explain L/T spectra involves radiative convection triggered by unstable carbon chemistry (CO/CH$_4$) in BD atmospheres. \cite{Tremblin_2019} and \cite{Phillips_2020} have shown this mechanism can reproduce the spectra and colours of L/T transition dwarfs. BD atmospheres can most likely be affected by both convection and clouds. However, the observed increase in photometric and spectroscopic variability of BDs in the L/T transition (e.g., \citealt{Artigau2009}; Fig. 8 of  \citealt{Radigan_2014}) has long been considered strong evidence for the latter scenario (clouds). In this framework, BDs transitioning from dusty to clear atmospheres as they cool would exhibit inhomogeneous cloud patterns, causing an increased variability until the entire cloud deck descends below the photosphere. \cite{Tremblin_2020} argue that for the sample of variable L/T BDs they studied, spectral modulation modelling proved degenerate between using cloud opacity or temperature variations due to convection.

Of the atmospheric models considered in this work, only the BT-Settl grids of \cite{Allard_2012} attempt to reproduce the features of the L/T transition with cloudy atmospheres. Although the grid usually performs well at reproducing the general trend of L/T dwarfs observed with photometry, such as the Spitzer data used in this work, our analysis shows that BT-Settl is unable to model the WFC3 spectroscopic features of LHS~6343\,C at a similar level of performance that is obtained using chemical non-equilibrium models. Although this work's results do not incorporate more recent cloud models (e.g., Exo-REM of \citealt{Charnay_2018}, Sonora-Diamondback of \citealt{morley2024sonora}), the fact that the ATMO-2020 NEQ models were the only ones successful in reproducing the entire available LHS~6343\,C spectrum, while predicting physical parameters consistent with Gaia and/or other photometric distances, does support the chemical non-equilibrium mechanism as a viable one to explain some L/T transition atmospheres. Nevertheless, a complete atmospheric and evolutionary analysis of LHS~6343\,C demands the use of both recent cloudy and chemical non-equilibrium BD models to explore the potential differences in the model fits and inferred physical parameters. In particular, the Spitzer photometry was insufficient to distinguish between cloudy or chemical non-equilibrium mechanisms. Modelling this bandpass with higher-resolution spectroscopy could help alleviate this degeneracy.

\vspace{0.3cm}
\subsection{\textbf{Viewing Angle}}
\label{subsec:viewingangle}
LHS~6343~C is both red in $J-H$ colour and faint in $J$-band magnitude with respect to the T0-T2 BDs distribution of Fig.~\ref{fig:CMD}. This may be evidence for a viewing geometry effect. Assuming that the recent finding that L3-L7 BDs seen equator-on show an excess color of $J-K\approx0.5$~mag \citep{suarez.2023, vos.2017} can be extrapolated to the L/T transition, then LHS~6343~C, most likely seen equator-on, should be expected to be both redder and fainter than the rest of the average field T0-T2 population. The effect is of order 0.5~mag in $J-K_s$ colours, equivalent to roughly 0.35~mag in $J-H$ (given that the average colours of T0-T2 are $J-K_s = 1.0$, $J-H=0.7$, \cite{leggett.2001}) with most of the attenuation afflicting the $J$-band, $\approx0.3$~mag. In other words, LHS~6343~C would be $\approx0.3$~mag brighter in $J$ and $\approx0.35$~mag bluer in $J-H$ if it were seen at the average viewing angle.

\vspace{0.3cm}
\subsection{\textbf{Host Star Effects on LHS~6343\,C }}

\cite{Montet_2016} had estimated the effect of irradiation on the BD's luminosity budget coming from LHS~6343\,A to be $\sim1\%$. Since this analysis used a different distance measure and determined a different $T_{\text{eff}}$, we re-perform this estimate to validate the assumption that LHS~6343\,C is minimally irradiated, and thus can adequately represent the properties of field BDs of its spectral type without the use of irradiation modelling. Using the LHS~6343\,A radius of Table~\ref{tab:TheBigOne} and the \cite{Mann_2015} radius-$T_{\text{eff}}$ empirical relation, we obtain an effective temperature for the M dwarf of $\sTEFFa$. Given these estimates and the known distance between the star and its companion BD, as well as assuming all incident flux is absorbed and re-emitted (i.e. Bond Albedo of 0), we calculate an equilibrium temperature for the BD of $\sTEQc$. Therefore, the emitted flux of the BD from absorption and re-emission of stellar host radiation is at most $\sBUDGET$ of its total luminosity budget, remaining negligible compared to the BD's measured luminosity.

In addition, Ohmic dissipation (\citet{Batygin_2010}, i.e., the conversion of electrical energy into heat due to the movement of charged particles in a magnetic field) could be transferring up to $10\%$ of incoming radiation to the interior depending on the strength of the BD's magnetic field \citep{Menou_2012}. Since LHS~6343\,C intercepts $3.75\pm0.74\times10^{19}$ W from its host star, Ohmic dissipation would at best provide an increase to the interior luminosity of $-8.02\pm0.08$ log($L_C/L_\odot$), which remains insufficient to meaningfully increase LHS~6343\,C's luminosity.

Significant tidal heating due to an ongoing circularization of the BD's orbit is also unlikely. LHS~6343\,C has a slightly elliptical orbit, so it has not fully circularized in its $\geq3$\,Gyr existence. We use the tidal heating rate equation of \cite{Jackson_2008} to determine a slight overestimate of the energy contributed by such a phenomenon as it applies to LHS~6343\,C. Using the physical parameters given in  Table~\ref{tab:TheBigOne}, we can make a rough lower estimate of the tidal quality factor $Q_{BD}=10^4$ informed from studies by \cite{beatty.2018} and \cite{Heller_2010}. We also use a generous upper estimate of the $k$ Love number for BDs $k_{2}=0.4$, informed from a study by \cite{Becker_2018} (Table 3). The resulting tidal heating rate would contribute only $-10.46\pm0.09$ log($L_T/L_\odot$) to the energy budget of LHS~6343\,C, which is several orders of magnitude lower than its measured luminosity. Other tidal phenomena of interest would be Kozai-Lidov cycles \citep{Kozai_1962,Lidov_1962} caused by the secondary M dwarf's orbit. \cite{Montet_2016} give an initial analysis of potential such cycles for the system, finding the timescales of the oscillations to likely be sufficiently smaller than the age of the system. However, a lack of astrometric observations on LHS~6343\,B makes it difficult to establish tighter constraints. Nevertheless, the contribution to tidal heating from this mechanism is likely insignificant compared to the BD's luminosity.

Finally, \cite{Montet_2016} also argue that any high energy radiation (via the stellar activity of the host M dwarf) that may have once influenced the atmosphere of LHS~6343\,C has likely been at low levels for billions of years, further allowing it to achieve an equilibrium representative of field BDs. Furthermore, the best outcomes of constrained atmospheric and evolutionary model fitting of Sections~\ref{subsubsec:constrainModFit} and \ref{subsec:evol} do not indicate a blatant disagreement between models and observations. All the points discussed here serve to further support the notion that LHS~6343\,C can be considered a valid analog to an isolated field BD of the same spectral type.

\section{\textbf{Conclusions}}

This work presents an original analysis of a secondary eclipse of the brown dwarf LHS~6343\,C using the HST/WFC3 IR grism.  We make use of prior work from \cite{Montet_2015,Montet_2016} presenting an analysis of Kepler transit, Spitzer/IRAC secondary eclipse and Keck/HIRES radial velocity observations. We update the brown dwarf (BD) mass and radius reported by Montet et al. using the empirical stellar mass and radius relations of \cite{Mann_2019} and \cite{Mann_2015}, as well as a Gaia DR3 distance of $\sDIST$, yielding $\sMASSc$ and $\sRADc$. 

Our WFC3 spectrum reveals a $\sSPTc$ spectral type for LHS~6343\,C. Its position in a $J$ vs. $J$\,$-$\,$H$ color-magnitude diagram indicates that this BD lies well within the L/T transition, although slightly redder and fainter than other T0-T2 dwarfs. This could be explained by its equator-on viewing geometry, assuming that the trend in reddening due to dust absorption seen in L dwarfs extends into the L/T transition \citep{suarez.2023}. A slight underestimate of the Gaia distance is also a possible explanation.

When combined with Kepler and Spitzer photometry from Montet et al., we obtain the most complete spectral energy distribution (SED) currently available for this BD. Gaps in the wavelength coverage of observations are filled using the best-fitting atmospheric model determined in section~\ref{subsubsec:unconstrainModFit}. Integrating this SED and adopting the Gaia distance and BD radius (which depends on distance) yields a bolometric luminosity of log($L_{\text{bol}}/L_{\sun}$) = $\vLBOLc \pm \eLBOLc$, of which $L_{\text{bol}}$ has $\sim$40\% of its total value and $\sim50\%$ of its total variance contributed by modelled bandpasses. Using the Stefan-Boltzmann law, we calculate an effective temperature $T_{\text{eff}}=\sTEFFc$; roughly $\sim100$-$200$~K warmer than previous estimates of \cite{Montet_2016} inferred only from Spitzer eclipses.

Finally, we use the BD's measured mass and luminosity (again dependant on Gaia) to interpolate both the ATMO-2020 and Sonora-Bobcat evolutionary models, yielding model-dependent ages of $\sAGEc$ and $\sAGEcSB$, respectively. Other predicted evolutionary parameters (e.g., radius) remained 1-$\sigma$ consistent with their measured counterparts obtained using the Gaia measurement.

Within this work, atmospheric characterization of LHS~6343\,C's observed SED was performed using the ATMO-2020, Sonora-Bobcat and BT-Settl (CIFIST 2011) models. Two approaches to modelling were considered. The first fixed the distance to the LHS~6343 system with a normal distribution about the Gaia DR3 value, which in turn constrained the possible mass and radius values. However, this distance measurement could be incorrect, since the DR3 processing pipeline ignored stellar binarity when modelling the system's parallax. Thus, our second approach to atmospheric characterization left the distance mostly unconstrained with a uniform prior between 9 and 90~pc.

The distance-constrained fits reveal that ATMO-2020 models with strong chemical non-equilibrium (NEQ) provide the best fit at $T_{\text{eff}}\approx1300\pm30\,\KELVIN$, with consistent physical parameter predictions regardless of the set of observation data being modelled. ATMO-2020 and Sonora-Bobcat chemical equilibrium (CEQ) models require higher temperatures ($\sim1330\pm30\,\KELVIN$) to model the HST/WFC3 spectrum, and lower temperatures ($\sim1100\pm55\,\KELVIN$) to model the Spitzer photometry. This tension is unsurprising, as these types of models are not expected to adequately reproduce the relative CO/CH$_4$ abundances giving rise to absorption bands at the Spitzer wavelengths. The BT-Settl models are incapable of adequately fitting the HST/WFC3 spectrum, only providing a good fit for the Spitzer photometry with a $T_{\text{eff}}$ prediction that is $\sim100\,\KELVIN$ lower ($1180\pm25\,\KELVIN$) compared to ATMO-2020 NEQ models.

For fits where distance was given a large uniform prior, ATMO-2020 NEQ models offer again the better fits across all instrument datasets modelled. The fits to the full SED or only to the Spitzer data remained consistent within 1-$\sigma$ of physical parameters measured using the Gaia distance, although noticeably greater distances ($\sim45\pm15\,\PARSEC$) were preferred, which translated to greater mass and radius values. Note that the increased distances obtained from unconstrained substellar atmospheric modelling are more consistent with the photometric distance obtained from an L/T spectral type - magnitude empirical relation ($\sim50\pm10\,\PARSEC$), and much less so with a photometric distance obtained using a similar relation in the M dwarf regime ($\sim27\pm8\,\PARSEC$).


The models most indifferent to the presence of physical constraints during fitting were the non-equilibrium chemistry models of ATMO-2020, whose inferred $T_{\text{eff}}$ remained within error for all cases except when the data were subject to an unconstrained fit of the HST/WFC3 spectrum only. Additionally, no combination of physical constraints or reduced instrument data during a fit was able to reproduce the totality of the HST/WFC3 spectrum properly. Most notably, the depth of the $1.4~\mu m$ water absorption band could only be properly reproduced with models having $T_{\text{eff}}\sim1450$~K. Hopefully, this may be corrected with the latest generation of models that include cloudy and/or chemical non-equilibrium prescriptions; e.g. Sonora-Diamondback of \cite{morley2024sonora}, Sonora Elf-Owl of \cite{mukherjee2024sonora}.

The atmospheric modelling results discussed in sections \ref{subsubsec:constrainModFit} and \ref{subsubsec:unconstrainModFit} point to the importance of spectroscopic observations with wide wavelength coverage for adequately testing current brown dwarf models in the L/T transition regime. 
Future work on LHS~6343\,C should therefore focus on panchromatic observations (e.g., JWST NIRSpec and MIRI/LRS) to enable better testing of brown dwarf models at a higher resolution across multiple bandpasses. Better spectral characterization of the two LHS~6343 M dwarfs and a more robust distance measurement would also prove very beneficial.

\begin{acknowledgments}

This research used observations made with the NASA/ESA \textit{Hubble Space Telescope} (HST) (DOI: \dataset[10.17909/3rbf-vc70]{http://dx.doi.org/10.17909/3rbf-vc70}). The data were downloaded from the \textit{Mikulski Archive for Space Telescopes} (MAST) managed by the \textit{Space Telescope Science Institute} (STScI). Relevant pipeline calibration files for HST/WFC3 were also taken from STScI.

This research used data from the \textit{Kepler} and \textit{Spitzer} Space Telescopes (KST, SST). KST funding was provided by the NASA Science Mission Directorate. SST was operated by the Jet Propulsion Laboratory, California Institute of Technology under a contract with NASA.

This research used data from the European Space Agency (ESA) mission {\it Gaia} (\url{https://www.cosmos.esa.int/gaia}), processed by the {\it Gaia} Data Processing and Analysis Consortium (DPAC, \url{https://www.cosmos.esa.int/web/gaia/dpac/consortium}). Funding for the DPAC has been provided by national institutions, in particular the institutions participating in the {\it Gaia} Multilateral Agreement.

This research has benefited from the SpeX Prism Spectral Libraries, maintained by Adam Burgasser at \url{http://pono.ucsd.edu/~adam/browndwarfs/spexprism}.

This research has made use of the SIMBAD database, operated at CDS, Strasbourg, France. It also benefited from The UltracoolSheet at \url{http://bit.ly/UltracoolSheet}, maintained by Will Best, Trent Dupuy, Michael Liu, Aniket Sanghi, Rob Siverd, and Zhoujian Zhang. Additionally, this research made use of the Montreal Open Clusters and Associations (MOCA) database, operated at the Montréal Planétarium (J. Gagné et al., in preparation). Finally, this research made use of the VizieR catalogue access tool, CDS, Strasbourg, France (DOI : 10.26093/cds/vizier). The original description of the VizieR service was published in 2000, A\&AS 143, 23 .

R.~D. acknowledges the support of the Natural Sciences and Engineering Research Council of Canada (NSERC) and the Trottier Family Foundation.

\end{acknowledgments}

\facilities{HST, Kepler, Spitzer, Gaia}

\software{\texttt{Astropy} \citep{astropy_2022},
          \texttt{emcee} \citep{Foreman_Mackey_2013},
          \texttt{corner} \citep{Foreman-Mackey2016},
          \texttt{BATMAN} \citep{Kreidberg_2015},
          \texttt{SPLAT} \citep{burgasser2017spex},
          \texttt{uncertainties} (Eric O. LEBIGOT, http://pythonhosted.org/uncertainties/),
          \texttt{Matplotlib} \citep{Hunter:2007},
          \texttt{NumPy} \citep{harris2020array},
          \texttt{SciPy} \citep{2020SciPy-NMeth},
          \texttt{ExoTEP} \citep{Benneke_2019v1,Benneke_2019v2}
}




This appendix presents an additional figure relevant to the HST WFC3 light curve fitting performed in section~\ref{sec:Datanal}. It shows the individual spectroscopic light curve data, as well as their respective best-fitting \texttt{BATMAN} models.

\vspace{0.0cm}

\begin{figure}[!hb]
\centering
\includegraphics[scale=0.275]{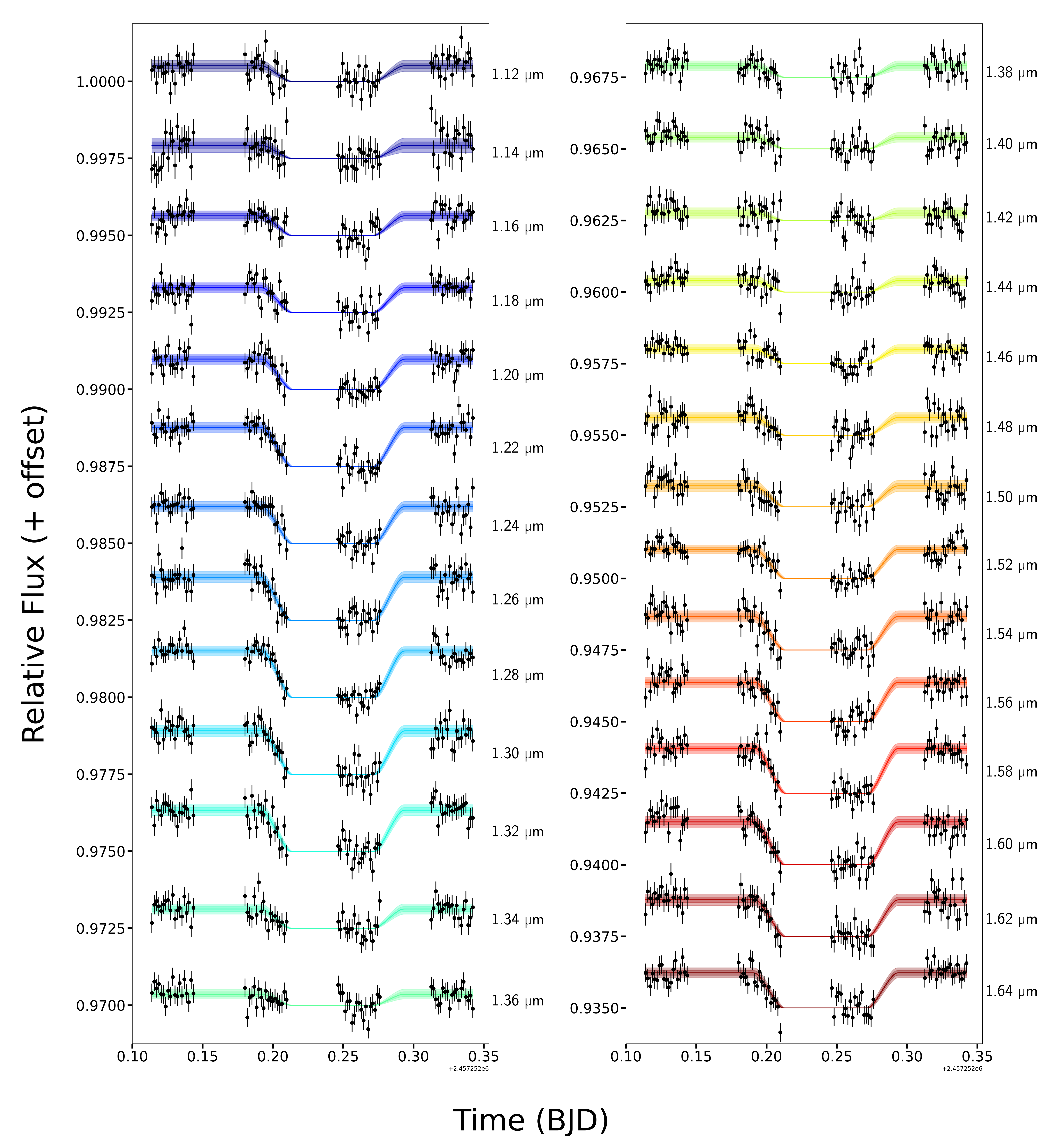}
\caption{The HST/WFC3 relative spectroscopic light curves of the observed LHS~6343\,C eclipse, corrected for instrument systematics. Each curve shows the median best-fit model as a full line and the $1\sigma$ and $2\sigma$ confidence intervals in lighter shades.}
\label{fig:SLC}
\end{figure}

\bibliography{article_refs}
\bibliographystyle{aasjournal}





\end{document}

%% file: LHS_6343_properties.tex
\newcommand{\lbolSUN}{L_{\odot}}
\newcommand{\massSUN}{M_{\odot}}
\newcommand{\radiusSUN}{R_{\odot}}
\newcommand{\massJUP}{M_{\text{Jup}}}
\newcommand{\radiusJUP}{R_{\text{Jup}}}
\newcommand{\KELVIN}{{\rm K}}
\newcommand{\PARSEC}{{\rm pc}}
\newcommand{\Gyr}{{\rm Gyr}}


\newcommand{\vDIST}{35.67}
\newcommand{\eDIST}{1.77}
\newcommand{\uDIST}{\PARSEC}
\newcommand{\sDIST}{\vDIST\pm\eDIST\,\,\uDIST}
\newcommand{\vDISTg}{35.67}
\newcommand{\eDISTg}{0.59}
\newcommand{\uDISTg}{\PARSEC}
\newcommand{\sDISTg}{\vDISTg\pm\eDISTg\,\,\uDISTg}
\newcommand{\vFeH}{0.03}
\newcommand{\eFeH}{0.26}
\newcommand{\uFeH}{{\rm [Fe/H]}}
\newcommand{\sFeH}{\vFeH\pm\eFeH\,\,\uFeH}
\newcommand{\vMETCON}{0.02}
\newcommand{\eMETCON}{0.19}
\newcommand{\uMETCON}{{\rm [a/H]}}
\newcommand{\sMETCON}{\vMETCON\pm\eMETCON\,\,\uMETCON}

\newcommand{\vKp}{13.104}
\newcommand{\eKp}{0.04}
\newcommand{\nKp}{$\textit{K}_{\hspace{0.01cm}P\hspace{0.02cm},\hspace{0.07cm}A+B}$}
\newcommand{\sKp}{\vKp\pm\eKp}
\newcommand{\vKpBA}{0.84}
\newcommand{\eKpBA}{0.12}
\newcommand{\nKpBA}{$\Delta\textit{K}_{\hspace{0.04cm}P\hspace{0.02cm},\hspace{0.07cm}B-A}$}
\newcommand{\sKpBA}{\vKpBA\pm\eKpBA}
\newcommand{\vKs}{8.695}
\newcommand{\eKs}{0.011}
\newcommand{\nKs}{$\textit{K}_{\hspace{0.01cm}S\hspace{0.02cm},\hspace{0.07cm}A+B}$}
\newcommand{\sKs}{\vKs\pm\eKs}
\newcommand{\vKsBA}{0.45}
\newcommand{\eKsBA}{0.06}
\newcommand{\nKsBA}{$\Delta\textit{K}_{\hspace{0.04cm}S\hspace{0.02cm},\hspace{0.07cm}B-A}$}
\newcommand{\sKsBA}{\vKsBA\pm\eKsBA}

\newcommand{\vMASSa}{0.363} 
\newcommand{\eMASSa}{0.020} 
\newcommand{\uMASSa}{\massSUN}
\newcommand{\sMASSa}{\vMASSa\pm\eMASSa\,\,\uMASSa}
\newcommand{\vRADa}{0.375}
\newcommand{\eRADa}{0.019}
\newcommand{\uRADa}{\radiusSUN}
\newcommand{\sRADa}{\vRADa\pm\eRADa\,\,\uRADa}
\newcommand{\vTEFFa}{3432}
\newcommand{\eTEFFa}{111}
\newcommand{\uTEFFa}{\KELVIN}
\newcommand{\nTEFFa}{T_{\text{eff}\hspace{0.03cm},\hspace{0.07cm}A}}
\newcommand{\sTEFFa}{\vTEFFa\pm\eTEFFa\,\,\uTEFFa}
\newcommand{\vMASSb}{0.297} 
\newcommand{\eMASSb}{0.019} 
\newcommand{\uMASSb}{\massSUN}
\newcommand{\sMASSb}{\vMASSb\pm\eMASSb\,\,\uMASSb}
\newcommand{\vRADb}{0.318}
\newcommand{\eRADb}{0.018}
\newcommand{\uRADb}{\radiusSUN}
\newcommand{\sRADb}{\vRADb\pm\eRADb\,\,\uRADb}
\newcommand{\vTEFFb}{3328}
\newcommand{\eTEFFb}{97}
\newcommand{\uTEFFb}{\KELVIN}
\newcommand{\nTEFFb}{T_{\text{eff}\hspace{0.03cm},\hspace{0.07cm}B}}
\newcommand{\sTEFFb}{\vTEFFb\pm\eTEFFb\,\,\uTEFFb}

\newcommand{\vMASSc}{62.6} 
\newcommand{\eMASSc}{2.2}  
\newcommand{\uMASSc}{\massJUP}
\newcommand{\sMASSc}{\vMASSc\pm\eMASSc\,\,\uMASSc}
\newcommand{\vRADc}{0.788}
\newcommand{\eRADc}{0.043}
\newcommand{\uRADc}{\radiusJUP}
\newcommand{\sRADc}{\vRADc\pm\eRADc\,\,\uRADc}
\newcommand{\vTEFFc}{1303} 
\newcommand{\eTEFFc}{29}   
\newcommand{\uTEFFc}{\KELVIN}
\newcommand{\nTEFFc}{T_{\text{eff}\hspace{0.03cm},\hspace{0.07cm}C}}
\newcommand{\sTEFFc}{\vTEFFc\pm\eTEFFc\,\,\uTEFFc}
\newcommand{\vLBOLc}{-4.77} 
\newcommand{\eLBOLc}{0.03}  
\newcommand{\uLBOLc}{\text{log}(\lbolSUN)}
\newcommand{\sLBOLc}{\vLBOLc\pm\eLBOLc\,\,\uLBOLc}
\newcommand{\vAGEc}{2.86}  
\newcommand{\eAGEc}{\text{}_{-0.33}^{+0.40}}  
\newcommand{\uAGEc}{\Gyr}
\newcommand{\sAGEc}{\vAGEc\hspace{0.08cm}\eAGEc\,\,\uAGEc}

\newcommand{\vBBRc}{1.64}  
\newcommand{\eBBRc}{0.14}  
\newcommand{\uBBRc}{{\rm erg~s^{-1}~cm^{-2}}}
\newcommand{\sBBRc}{(\vBBRc\pm\eBBRc)\hspace{0.1cm}\times10^8\,\,\uBBRc}
\newcommand{\vAGEcSB}{3.11}  
\newcommand{\eAGEcSB}{\text{}_{-0.38}^{+0.50}} 
\newcommand{\uAGEcSB}{\Gyr}
\newcommand{\sAGEcSB}{\vAGEcSB\hspace{0.08cm}\eAGEcSB\,\,\uAGEcSB}
\newcommand{\vSPTc}{T1.5}
\newcommand{\eSPTc}{1}
\newcommand{\sSPTc}{\text{\vSPTc}\pm\text{\eSPTc}}
\newcommand{\vKsC}{14.87}
\newcommand{\eKsC}{0.11}
\newcommand{\nKsC}{$M_J$}
\newcommand{\sKsC}{\vKsC\pm\eKsC}
\newcommand{\vRHOc}{161}
\newcommand{\eRHOc}{23}
\newcommand{\uRHOc}{{\rm g~cm^{-3}}}
\newcommand{\sRHOc}{\vRHOc\pm\eRHOc\,\,\uRHOc}
\newcommand{\vLOGGc}{5.40}
\newcommand{\eLOGGc}{0.04}
\newcommand{\uLOGGc}{{\rm \text{log}(cm/s^2)}}
\newcommand{\sLOGGc}{\vLOGGc\pm\eLOGGc\,\,\uLOGGc}
\newcommand{\vSCALc}{2.62}
\newcommand{\eSCALc}{0.21}
\newcommand{\sSCALc}{(\vSCALc\pm\eSCALc)\times10^{-21}}
\newcommand{\vSMAc}{0.08}
\newcommand{\eSMAc}{0.004}
\newcommand{\uSMAc}{{\rm AU}}
\newcommand{\sSMAc}{\vSMAc\pm\eSMAc\,\,\uSMAc}
\newcommand{\vTEQc}{358}
\newcommand{\eTEQc}{12}
\newcommand{\uTEQc}{\KELVIN}
\newcommand{\sTEQc}{\vTEQc\pm\eTEQc\,\,\uTEQc}

\newcommand{\vBUDGET}{0.57}
\newcommand{\eBUDGET}{0.08}
\newcommand{\uBUDGET}{\%}
\newcommand{\sBUDGET}{\vBUDGET\pm\eBUDGET\,\,\uBUDGET}
\newcommand{\vINT}{-7.02}
\newcommand{\eINT}{0.08}
\newcommand{\uINT}{\text{log}(\lbolSUN)}
\newcommand{\sINT}{\vINT\pm\eINT\,\,\uINT}
\newcommand{\vTIDE}{-10.46}
\newcommand{\eTIDE}{0.09}
\newcommand{\uTIDE}{\text{log}(\lbolSUN)}
\newcommand{\sTIDE}{\vTIDE\pm\eTIDE\,\,\uTIDE}

%% file: affiliations.tex
\newcommand{\UdeM}{Département de Physique, Université de Montréal, C.P. 6128, Succ. Centre-ville, Montréal, H3C 3J7, Québec, Canada}

\newcommand{\iREx}{Trottier Institute for Research on Exoplanets, Université de Montréal}

\newcommand{\PRTA}{Planétarium Rio Tinto Alcan, Espace pour la Vie, 4801 Av. Pierre-de Coubertin, Montréal, Québec, Canada}

\newcommand{\HSCA}{Harvard-Smithsonian Center for Astrophysics, 60 Garden Street, Cambridge, MA 02138 USA}

\newcommand{\UNSW}{School of Physics, University of New South Wales, Sydney, NSW 2052, Australia}

\newcommand{\Udash}{UNSW Data Science Hub, University of New South Wales, Sydney, NSW 2052, Australia}

\newcommand{\SRON}{SRON, Netherlands Institute for Space Research, Niels Bohrweg 4, NL-2333 CA, Leiden, The Netherlands}